\documentclass[11pt,twoside]{article}
\usepackage{etex,amsmath, amssymb, amsthm, mathrsfs, chemarrow, amsfonts, bbm, ctable,graphicx,color,epsfig,citesort, stmaryrd,textpos, upgreek, xy, extarrows,protosem,fancybox}

\usepackage[text={6.9in,9in},centering,includehead,includefoot,headheight=14pt]{geometry}
\usepackage[version=3,arrows=pgf-filled]{mhchem} 
\usepackage[all,cmtip]{xypic}

\usepackage[abs]{overpic}

\usepackage{fancyhdr}
\renewcommand{\subsectionmark}[1]{}
\newlength\Li \newlength\Lii
\date{}

\makeatletter
\def\@seccntformat#1{}
\makeatother
\numberwithin{equation}{section}
\renewcommand{\numberline}[1]{}

\makeatletter
\def\blfootnote{\xdef\@thefnmark{}\@footnotetext}
\makeatother

\title{Dynamic $p$-enrichment schemes for multicomponent reactive flows} 

\author{C.~Michoski\textsuperscript{\htmladdnormallink{\textproto{\AAsade}}{https://webspace.utexas.edu/michoski/Michoski.html}} , \ C.~Mirabito,  \ C.~Dawson, \\  \small{ \it Institute for Computational Engineering and Sciences (ICES), Computational Hydraulics Group (CHG)} \\ \small{\it University of Texas at Austin, Austin, TX, 78712} \\ \\ E.~J.~Kubatko,  \\  \small{ \it Department of Civil and Environmental Engineering and Geodetic Sciences}  \\ \small{\it The Ohio State University, Columbus, OH, 43210} \\ \\  D.~Wirasaet, \ J.~J.~Westerink \\ \small{\it Computational Hydraulics Laboratory, Department of Civil Engineering and Geological Sciences} \\ \small{\it University of Notre Dame, Notre Dame, IN, 46556}}

\usepackage[square,comma,numbers,sort&compress]{natbib}
\usepackage[pdftex,bookmarks]{hyperref}
\usepackage{hypernat}

\begin{document}
\maketitle
\begin{abstract}  

We present a family of $p$-enrichment schemes.  These schemes may be separated into two basic classes: the first, called \emph{fixed tolerance schemes}, rely on setting global scalar tolerances on the local regularity of the solution, and the second, called \emph{dioristic schemes}, rely on time-evolving bounds on the local variation in the solution.  Each class of $p$-enrichment scheme is further divided into two basic types.  The first type (the Type I schemes) enrich along lines of maximal variation, striving to enhance stable solutions in ``areas of highest interest.''  The second type (the Type II schemes) enrich along lines of maximal regularity in order to maximize the stability of the enrichment process.  Each of these schemes are tested over a pair of model problems arising in coastal hydrology.  The first is a contaminant transport model, which addresses a declinature problem for a contaminant plume with respect to a bay inlet setting.   The second is a multicomponent chemically reactive flow model of estuary eutrophication arising in the Gulf of Mexico.\\ \\ \small{{\bf Keywords}: $p$-adaptivity, $p$-enrichment, discontinuous Galerkin, finite elements, RKDG, RKSSP, dynamic $p$-enrichment, flow reactors, advective transport, shallow water equations, contaminant transport, contaminant declinature, estuary eutrophication.}

\end{abstract}

\blfootnote{\textsuperscript{\htmladdnormallink{\textproto{\AAsade}}{https://webspace.utexas.edu/michoski/Michoski.html}}Corresponding author, {\it michoski@ices.utexas.edu}}

\tableofcontents

\section{\texorpdfstring{\protect\centering $\S 1$ Introduction}{\S 1  Introduction}}

The topic of $p$-enrichment holds substantial allure in computational engineering and numerics due to the promise of being able to locally resolve solution structure without paying the full cost of raising the global order of the solution a complete integral step.  Though a good number of important results exist regarding $p$-enrichment, the majority of these results deal with $p$-adaptation as the fundamentally coupled component of an $hp$-adaptive scheme, wherein the numerical behavior of the $p$-enrichment is not isolated from the behavior of the $h$-adaptivity.  Moreover, of the work that does address $p$-enrichment as its own topic, there is often the strong presence of assumptions restricting the resulting behaviors to highly idealized linear systems of static equations, or to otherwise highly specialized regimes aimed at very particular applications and model systems. In this paper we attempt to abate this absence by addressing the merits and limitations of a family of $p$-enrichment schemes implemented in a discontinuous Galerkin formulation to a generalized system of multicomponent equations.

A brief note should be made here about our lexemic designation.  That is, much of the literature on $p$-enrichment refer to it merely as $p$-adaptivity.  This takes on a clear meaning, particularly in the context of $hp$-adaptivity, where it generally describes what is happening in that context, in that $p$ ``adapts'' along the lines of $h$ in an often very complicated and highly $hp$-coupled manner.  However, in the restricted case where one is only $p$-adapting, the context that emerges is a bit less subtle.  Here, the general idea is to take some solution order $p$ (where $p=1$ would correspond to linears), and attempt to raise the global average order of the solution without having to spend the computational resources required to raise the order of the solution in every element (\emph{e.g.} $p=2$ globally).  Taken in this context it is then natural to view the ultimate goal of a purely $p$-adaptive scheme to be that of: global $p$-enrichment. Some authors however also prefer to use the term `$p$-refinement' instead of `$p$-enrichment.'  Though this usage seems fairly clear in meaning, we prefer `$p$-enrichment' to `$p$-refinement' here, where the concept of \emph{refinement} and \emph{coarsening} seems naturally applicable to the mesh operations that occur in $h$-adaptivity, while in the context of $p$-adaptivity we reserve \emph{enrichment} and \emph{de-enrichment} for the contrasting operations applied with respect to the order of the solution.  Finally, it should be noted that P-enrichment may refer in chemical, biological and ecological settings to a type of phosphorous-enrichment, and should not be confused with $p$-enrichment in the numerical sense used in this paper, albeit in \textsection{4} to chemical problems involving the formation and depletion of aqueous phosphate levels.

Within the context of $hp$-adaptivity, the two volume work of Demkowicz,~\emph{et.~al.} \cite{Dem,Demk2} provides both a nice introduction to some of the basic emergent features of $hp$-adaptive regimes, as well as some advanced topics, such as $hp$-adaptive schemes on boundary value problems applied to (a time-independent form of) Maxwell's equations; however, it should be noted that little emphasis is made specifically to discontinuous Galerkin (DG) formulations.  Nevertheless, many of the basic principles that underlie $hp$-adaptive methodologies can be viewed as relatively method--independent.  For $hp$-adaptive schemes applied specifically to DG formulations, the work of \cite{Schotzau} applies an $hp$-adaptive scheme to systems of nonlinear ODEs where an emphasis is made on exact numerical and mathematical behaviors of well-behaved solutions that demonstrate exponential rates of convergence in time. In \cite{Dedner} a nice $hp$-adaptive scheme is presented which is based on using sharp \emph{a posteriori} error estimates over (non)linear advective PDEs.  In \cite{Bey1,Bey2} a parallel $hp$-adaptive method is developed that uses \emph{a posteriori} error tolerances in order to enforce sharp bounds on the error convergence.  The $hp$-adaptive scheme presented in \cite{WanMa} uses a pair of ``smoothness indicators'' over either the interior of the cell for $p$-enrichment, or with respect to the jump discontinuity between cells for $h$-adaptation. In \cite{dem2} an energy norm approach is given for Burger's and the Navier-Stokes equations, where artificial viscosity is added in the case of Burger's equations in order to develop the ``diffusion scale'' which serves as the implicit tolerance by which optimal convergence in norm is achieved.  Additionally, we have recently developed an $hp$-adaptive energy method in \cite{MES} based on \emph{a priori} stability results, which use the local \emph{entropy} of the fully coupled system of equations to ``sense'' and preserve global energy bounds, that we then apply to large multicomponent systems of reaction--diffusion equations.

The results relating to $p$-enrichment, on the other hand, are more sparse.  In the context of discontinuous Galerkin systems the two most prevalent results can be found in \cite{BurbeauS,KubBDW}, which are discussed in detail in \textsection{3}.  Briefly, in \cite{BurbeauS} Burbeau and Sagaut develop what they refer to as ``regularity sensors'' for $p$-enrichment, and apply this model to Euler's equations in one and two dimensions from an applied engineering perspective.  Next, the work of Kubatko, \emph{et al}. in \cite{KubBDW} expands on the model from \cite{BurbeauS}, where a similar type of $p$-enrichment method is employed, but the $L^{1}$-error behavior is carefully analyzed for dynamic $p$-enrichment with respect to varying fixed $h$-refinement levels.  In this work, both of these schemes would fit into the category of \emph{Type I fixed tolerance schemes}, and both will be discussed in detail below in \textsection{3}.

We begin by introducing the generalized system of equations that we employ in \textsection{2}, where we couch this system of initial-boundary value partial differential equations in the setting of a discontinuous Galerkin finite element method.  Then in \textsection{3} we introduce and discuss a number of different general types of $p$-enrichment schemes, which include two types of both \emph{fixed tolerance} as well as \emph{dioristic schemes}.

Finally, let us briefly discuss some of the engineering applications that we are interested in.  Our generalized system of equations (see \ref{system}) encases a very large class of possible applications, but here we restrict ourselves to a pair of example applications arising in the field of coastal hydrology.    In this setting, it has long been recognized that the shallow water equations offer a good and soluble approximation to flow regimes present in complicated physical systems (\emph{e.g.} in coastal flows and storm surge models \cite{BD2,Vre,Dawson2010}, \emph{etc.}).  Here the depth--averaged shallow water equations (SWE) provide solutions to the elevation of the free surface of the water $\zeta=\zeta(t,\boldsymbol{x})$ and the velocity component $\boldsymbol{u}=\boldsymbol{u}(t,\boldsymbol{x})$ of the flow as a fully coupled system of partial differential equations.   Since this formulation models the essential mechanical properties of the flow pointwise as a function of time, it is clear that an advected scalar quantity (as we call $\iota$ in \textsection{4}) satisfying the linear transport equation \begin{equation}\label{transport}\iota_{t}+\boldsymbol{u}\cdot\nabla_{x}\iota = 0, \end{equation} coupled to the SWEs must track the flow characteristics of the solution, and thus for any chemically inert constituent $\iota$ of the flow field represents the \emph{field of dispersal} of that inert constituent, up to the depth--averaged approximation.  Such a representation is particularly compelling when the constituent $\iota$ is reasonably well-mixed (homogeneous) across the vertical stratification of the density field.  Moreover, in such a case, it immediately follows that a chemical mass action principle must also be satisfied at the same corresponding scales.  

We present two separate models in \textsection{4} that employ this basic assumption in the context of coastal hydrology and the SWEs, and which attempt to take a first step towards addressing a pair of important questions concerning environmental coastal remediation.  The form of the contaminant transport equation in \textsection{4.1} can be easily derived from first principles, though one should refer to \cite{Heemink,Ors,Ors2,Aizinger1,Aizinger2} for background on its uses and applications, and one should further note its close mathematical likeness to sediment transport models \cite{Mirabito}.  One thing to note is that we have suppressed the Fickian diffusion often present in the contaminant transport (\ref{transport}) aspects of the model.  We have chosen this slight simplification for this paper in part due to a frequent criticism leveled \cite{LipRing} against Eulerian frame solutions to the transport of passive tracers in contaminant models, which points out that numerical diffusion can become excessive for finite element solutions that are less than second order accurate in space. In fact, this provides one of the basic motivating factors for solutions that implement Lagrangian frame tracer particles in fixed lower order spatial schemes \cite{Ors2,Luettich}, which as a consequence demonstrate no diffusion at all (neither numerical nor Fickian), but rather travel indefinitely along the characteristic field.  Since there remains a question as to what the role and relationship between the numeric dampening and the Fickian diffusion in the setting of a $p$-adaptive regime should be, where the polynomial order may span (by construction) the set $p\in\{0,1,\ldots,n\}$, we save this topic of macroscopic particle diffusion for another time (though we also refer the reader to \cite{MES} for some preliminary results in this direction).

Our second example in \textsection{4.2} takes the standard transport model from \textsection{4.1} and extends it to include reactive chemical constituents.  We do this by coupling a standard chemical mass action term.  This type of reactive shallow water model has been implemented recently in \cite{Alvarez,Cheng}, and can be inferred from many related systems, such as \cite{MacQuarrie,Aizinger1}.  We provide a fully Eulerian solution to the multicomponent reactive system of equations, by implementing the discontinuous Galerkin model for chemical reactor systems developed in \cite{MES}, and apply this system to a realistic mesh of the Brazos estuary emptying into the Gulf of Mexico, and analyzed in the context of the development of contaminant induced  ``dead zones'' in these types of regions.

\section{\texorpdfstring{\protect\centering $\S 2$ Governing equations and formulation}{\S 2 Governing equations and formulation}}

We are primarily focused in this paper on systems related to the shallow water equations \cite{BD2}, though in this setting we wish to include systems that have both complicated reaction dynamics and free boundary data.  Thus, we generalize somewhat to consider the following initial--boundary value problem in $\Omega\times (0,T)$, taking $\Omega\subset\mathbb{R}^{2}$ with boundary $\partial\Omega$, such that the system is governed by: \begin{equation}\label{system}\boldsymbol{U}_{t} + \boldsymbol{F}_{x} - \boldsymbol{G}_{x} = \boldsymbol{g}, \quad \mathrm{given \ initial \ conditions} \quad\boldsymbol{U}_{|t=0}=\boldsymbol{U}_{0},\end{equation} with Robin free boundary, \begin{equation} \label{bounds}a_{i}U_{i} + \nabla_{x}U_{i,x}\left( b_{i}\cdot \boldsymbol{n} + c_{i}\cdot \boldsymbol{\tau}\right) - f_{i} = 0,\quad \mathrm{on} \ \partial\Omega_{0}.\end{equation} 

More precisely, the system is comprised of an $m$--dimensional state vector $\boldsymbol{U}=\boldsymbol{U}(t,\boldsymbol{x}) = (U_{1},\ldots,U_{m})$, an advective flux matrix $\boldsymbol{F} = \boldsymbol{F}(\boldsymbol{U})$, a viscous flux matrix $\boldsymbol{G}=\boldsymbol{G}(\boldsymbol{U},\boldsymbol{U}_{x})$, and a source term $\boldsymbol{g}=\boldsymbol{g}(t,\boldsymbol{x})= (g_{1},\ldots,g_{m})$, where $\boldsymbol{x}\in\mathbb{R}^{2}$ and $t\in(0,T)$.  The vectors $\boldsymbol{a}$, $\boldsymbol{b}$, $\boldsymbol{c}$ and $\boldsymbol{f}$ are comprised of the $m$ functions,  $a_{i}=a_{i}(t,\boldsymbol{x})$, $b_{i}=b_{i}(t,\boldsymbol{x})$, $c_{i}=c_{i}(t,\boldsymbol{x})$ and $f_{i}=f_{i}(t,\boldsymbol{x})$ for $i=1,\ldots,m$, where $\boldsymbol{n}$ denotes the unit outward pointing normal, and $\boldsymbol{\tau}$ the unit tangent vector.   The time--varying free boundary $\partial\Omega_{free}=\partial\Omega_{free}(t,\boldsymbol{x})$ adjoined to the absolute domain boundary $\partial\Omega$ make up what we refer to as ``the effective boundary,'' which we denote by $\partial\Omega_{0}=\partial\Omega\cup\partial\Omega_{free}$ (for example, we use a wetting and drying boundary condition below that corresponds to the free boundary).

In addition, because we are interested in approximate numerical solutions of the form of \cite{Aizinger1,ABCM} restricted in part to the family of LDG (\emph{local} discontinuous Galerkin) methods as developed originally for elliptic equations \cite{ABCM}, we rewrite (\ref{system}) as a coupled system in terms of an auxiliary variable $\boldsymbol{\Sigma}$, such that the system we solve becomes:  \begin{equation}\begin{aligned}\label{system2}\boldsymbol{U}_{t} + \boldsymbol{F}_{x} - \boldsymbol{G}_{x} = \boldsymbol{g}, \quad\mathrm{and}\quad \boldsymbol{\Sigma} = \boldsymbol{U}_{x},\end{aligned}\end{equation} where we have substituted the auxiliary variable into the viscous flux matrix, so that $\boldsymbol{G} = \boldsymbol{G}(\boldsymbol{U},\boldsymbol{\Sigma})$.

For notational completeness we adopt the following discretization scheme motivated by \cite{MESV,FFS}.  Consider the open set $\Omega\subset\mathbb{R}^{2}$ with boundary $\partial\Omega$, given $T>0$ such that $\mathcal{Q}_{T}=((0,T)\times\Omega)$. Let $\mathscr{T}_{h}$ denote the partition of the closure of the polygonal triangulation of $\Omega$, which we denote $\Omega_{h}$, into a finite number of polygonal elements denoted $\Omega_{e}$, such that  $\mathscr{T}_{h}= \{\Omega_{e_1},\Omega_{e_2}, \ldots,\Omega_{e_{ne}}\}$, for $ne\in\mathbb{N}$ the number of elements in $\Omega_{h}$.   In this work we define the mesh diameter $h$ to satisfy $h = \min_{ij}(d_{ij})$ for the distance function $d_{ij}= d(\boldsymbol{x}_{i},\boldsymbol{x}_{j})$ and elementwise face vertices $\boldsymbol{x}_{i},\boldsymbol{x}_{j}\in\partial\Omega_{e}$ when the mesh is structured and regular.  For unstructured meshes we mean the average value of $h$ over the mesh.

Now, let $\Gamma_{ij}$ denote the face shared by two neighboring elements $\Omega_{e_{i}}$ and $\Omega_{e_{j}}$, and for $i\in I\subset\mathbb{Z}^{+}=\{1,2,\ldots\}$ define the indexing set $r(i)=\{j \in I : \Omega_{e_{j}}$ is a neighbor of $\Omega_{e_{i}}\}$.  Let us denote all $\Omega_{e_{i}}$ containing the boundary $\partial\Omega_{h}$ by $S_{j}$ and letting $I_{B}\subset \mathbb{Z}^{-}=\{-1,-2,\ldots\}$ define $s(i)=\{j\in I_{B}:S_{j}$ is a face of $\Omega_{e_{i}}\}$ such that $\Gamma_{ij}=S_{j}$ for $\Omega_{e_{i}}\in \Omega_{h}$ when $S_{j}\in\partial\Omega_{e_{i}}$, $j\in I_{B}$.  Then for $\Xi_{i}=r(i)\cup s(i)$, we have \[\partial\Omega_{e_{i}}=\bigcup_{j\in \Xi(i)}\Gamma_{ij},\quad\mathrm{and}\quad \partial\Omega_{e_{i}}\cap\partial\Omega_{h} = \bigcup_{j\in s(i)}\Gamma_{ij}.\] 

We are interested in obtaining an approximate solution to $\boldsymbol{U}$ at time $t$ on the finite dimensional space of discontinuous piecewise polynomial functions over $\Omega$ restricted to $\mathscr{T}_{h}$, given as \[S_{h}^{p}(\Omega_{h},\mathscr{T}_{h})=\{v:v_{|\Omega_{e_{i}}}\in \mathscr{P}^{p}(\Omega_{e_{i}}) \ \ \forall\Omega_{e_{i}}\in\mathscr{T}_{h}\}\] for $\mathscr{P}^{p}(\Omega_{e_{i}})$ the space of degree $\leq p$ polynomials over $\Omega_{e_{i}}$.      

Choosing a set of degree $p$ polynomial basis functions $N_{\ell}\in\mathscr{P}^{p}(\Omega_{e_{i}})$ for $\ell =1,\ldots, n_{p}$ the corresponding degrees of freedom, we can denote the state vector at time $t$ over $\Omega_{e_{i}}$, by
\begin{equation}\label{shapefunctions}
\boldsymbol{U}_{hp}(t,\boldsymbol{x})=\sum_{\ell=1}^{n_{p}}\boldsymbol{U}_{\ell}^{i}(t)N^{i}_{\ell}(\boldsymbol{x}),\quad  \forall x\in\Omega_{e_{i}},
\end{equation}
 where the $N^{i}_{\ell}$'s are the finite element shape functions in the DG setting, and the $\boldsymbol{U}_{\ell}^{i}$'s correspond to the unknowns.   We characterize the finite dimensional test functions \[\boldsymbol{v}_{hp},\boldsymbol{w}_{hp}\in W^{k,q}(\Omega_{h},\mathscr{T}_{h}),\quad\mathrm{by}\quad\boldsymbol{v}_{hp}(x)=\sum_{\ell=1}^{n_{p}}\boldsymbol{v}_{\ell}^{i}N_{\ell}^{i}(x)\quad\mathrm{and}\quad\boldsymbol{w}_{hp}(x)=\sum_{\ell=1}^{n_{p}}\boldsymbol{w}_{\ell}^{i}N_{\ell}^{i}(x)\] where $\boldsymbol{v}_{\ell}^{i}$ and $\boldsymbol{w}_{\ell}^{i}$ are the coordinates in each $\Omega_{e_{i}}$, with the broken Sobolev space over the partition $\mathscr{T}_{h}$ defined by \[W^{k,q}(\Omega_{h},\mathscr{T}_{h})=\{\omega : \omega_{|\Omega_{e_{i}}}\in W^{k,q}(\Omega_{e_{i}}) \ \ \forall\Omega_{e_{i}}\in\mathscr{T}_{h}\}.\]   

Thus, for $\boldsymbol{U}$ a classical solution to (\ref{system2}), multiplying by $\boldsymbol{v}_{hp}$ or $\boldsymbol{w}_{hp}$ and integrating elementwise by parts yields the coupled system: \begin{equation}\begin{aligned}\label{approxsystem} & \frac{d}{dt}\int_{\Omega_{e_{i}}}\boldsymbol{U}\cdot\boldsymbol{v}_{hp}dx +  \int_{\Omega_{e_{i}}} (\boldsymbol{F}\cdot\boldsymbol{v}_{hp})_{x} dx - \int_{\Omega_{e_{i}}} \boldsymbol{F}:\boldsymbol{v}^{hp}_{x}dx  \\ & \qquad\qquad - \int_{\Omega_{e_{i}}} ( \boldsymbol{G}\cdot\boldsymbol{v}_{hp})_{x} dx  + \int_{\Omega_{e_{i}}}  \boldsymbol{G}:\boldsymbol{v}^{hp}_{x}dx  =  \int_{\Omega_{e_{i}}}  \boldsymbol{v}_{hp}\cdot\boldsymbol{g}dx, \\ & \int_{\Omega_{e_{i}}}\boldsymbol{\Sigma}\cdot \boldsymbol{w}_{hp} dx - \int_{\Omega_{e_{i}}}(\boldsymbol{U}\cdot\boldsymbol{w}_{hp})_{x} dx +\int_{\Omega_{e_{i}}}\boldsymbol{U}:\boldsymbol{w}^{hp}_{x}dx=0,\end{aligned}\end{equation} where $(:)$ denotes the scalar product.

Now, let $\boldsymbol{n}_{ij}$ be the unit outward normal to $\partial\Omega_{e_{i}}$ on $\Gamma_{ij}$, and let $v_{|\Gamma_{ij}}$ and  $v_{|\Gamma_{ji}}$ denote the values of $v$ on $\Gamma_{ij}$ considered from the interior and the exterior of $\Omega_{e_{i}}$, respectively.  Then by choosing componentwise approximations in (\ref{approxsystem}) by substituting in (\ref{shapefunctions}), we arrive with the approximate form of the first term of (\ref{approxsystem}) given by, \begin{equation}
\begin{aligned}\label{timeterm}
\frac{d}{dt}\int_{\Omega_{e_{i}}}\boldsymbol{U}_{hp}\cdot \boldsymbol{v}_{hp}dx  \approx  \frac{d}{dt}\int_{\Omega_{e_{i}}}\boldsymbol{U}\cdot\boldsymbol{v}_{hp}dx,
\end{aligned}
\end{equation}  the second term using an inviscid numerical flux $\Phi_{i}$, by
\begin{equation}
\begin{aligned}\label{invflux}
\tilde{\Phi}_{i}(\boldsymbol{U}_{hp}|_{\Gamma_{ij}},\boldsymbol{U}_{hp}|_{\Gamma_{ji}}, \boldsymbol{v}_{hp}) & = \sum_{j\in \Xi(i)}\int_{\Gamma_{ij}}\Phi(\boldsymbol{U}_{hp}|_{\Gamma_{ij}},\boldsymbol{U}_{hp}|_{\Gamma_{ji}},\boldsymbol{n}_{ij})\cdot\boldsymbol{v}_{hp}|_{\Gamma_{ij}} d\Xi \\ & \approx  \sum_{j\in \Xi(i)} \int_{\Gamma_{ij}}\sum_{l =1}^{2}(\boldsymbol{F})_{l}\cdot (n_{ij})_{l}\boldsymbol{v}_{hp}|_{\Gamma_{ij}}d\Xi,
\end{aligned}
\end{equation} and the third term in (\ref{approxsystem}) by,
\begin{equation}\label{third}
\Theta_{i}(\boldsymbol{U}_{hp},\boldsymbol{v}_{hp})= \int_{\Omega_{e_{i}}} \boldsymbol{F}_{hp}:\boldsymbol{v}^{hp}_{x} dx \approx \int_{\Omega_{e_{i}}} \boldsymbol{F} : \boldsymbol{v}^{hp}_{x} dx.
\end{equation}

Next we approximate the boundary viscous term of (\ref{approxsystem}) using a generalized viscous flux $\hat{\mathscr{G}}$ such that,
\begin{equation}
\begin{aligned}
\label{viscous}
\mathscr{G}_{i}(\boldsymbol{\Sigma}_{hp},\boldsymbol{U}_{hp},\boldsymbol{v}_{hp}) & = \sum_{j\in \Xi(i)}\int_{\Gamma_{ij}}\hat{\mathscr{G}}(\boldsymbol{\Sigma}_{hp}|_{\Gamma_{ij}},\boldsymbol{\Sigma}_{hp}|_{\Gamma_{ji}}, \boldsymbol{U}_{hp}|_{\Gamma_{ij}}, \boldsymbol{U}_{hp}|_{\Gamma_{ji}}, \boldsymbol{n}_{ij})\cdot\boldsymbol{v}_{hp}|_{\Gamma_{ij}} d\Xi \\ & \approx  \sum_{j\in\Xi(i)}\int_{\Gamma_{ij}}\sum_{l=1}^{N}(\boldsymbol{G})_{l}\cdot (n_{ij})_{l}\boldsymbol{v}_{hp}|_{\Gamma_{ij}}d\Xi,
\end{aligned}
\end{equation} 
 while the second viscous term is approximated by:
\begin{equation}\label{second}\mathscr{N}_{i}(\boldsymbol{\Sigma}_{hp},\boldsymbol{U}_{hp},\boldsymbol{v}_{hp})=\int_{\Omega_{e_{i}}}\boldsymbol{G}_{hp}:\boldsymbol{v}^{hp}_{x}dx\approx \int_{\Omega_{e_{i}}}\boldsymbol{G}:\boldsymbol{v}^{hp}_{x}dx.\end{equation}

For the auxiliary equation in (\ref{approxsystem}) we expand it such that the approximate solution satisfies: \begin{equation} \begin{aligned}\label{penalty} \mathscr{Q}_{i}(\hat{\boldsymbol{U}},\boldsymbol{\Sigma}_{hp},\boldsymbol{U}_{hp},\boldsymbol{w}_{hp},\boldsymbol{w}_{x}^{hp}) & =\int_{\Omega_{e_{i}}} \boldsymbol{\Sigma}_{hp}\cdot \boldsymbol{w}_{hp}dx  + \int_{\Omega_{e_{i}}}\boldsymbol{U}_{hp} : \boldsymbol{w}^{hp}_{x}dx \\ & - \sum_{j\in \Xi(i)}\int_{\Gamma_{ij}}\hat{\boldsymbol{U}}(\boldsymbol{U}_{hp}|_{\Gamma_{ij}},\boldsymbol{U}_{hp}|_{\Gamma_{ji}},\boldsymbol{w}_{hp}|_{\Gamma_{ij}},\boldsymbol{n}_{ij}) d\Xi,\end{aligned}\end{equation} where, \[\begin{aligned} \sum_{i\in I}\sum_{j\in \Xi(i)}\int_{\Gamma_{ij}}\hat{\boldsymbol{U}}(\boldsymbol{U}_{hp}|_{\Gamma_{ij}},\boldsymbol{U}_{hp}|_{\Gamma_{ji}},\boldsymbol{w}_{hp}|_{\Gamma_{ij}},\boldsymbol{n}_{ij}) d\Xi \approx \sum_{i\in I}\sum_{j\in \Xi(i)}\int_{\Gamma_{ij}}\sum_{l=1t}^{N}(\boldsymbol{U})_{l}\cdot (n_{ij})_{l} \boldsymbol{w}_{hp}|_{\Gamma_{ij}}d\Xi\end{aligned}\] given a generalized numerical flux $\hat{\boldsymbol{U}}$, and where \[\int_{\Omega_{e_{i}}} \boldsymbol{\Sigma}_{hp}\cdot \boldsymbol{w}_{hp}dx \approx \int_{\Omega_{e_{i}}} \boldsymbol{\Sigma}\cdot \boldsymbol{w}_{hp}dx, \quad\mathrm{and}\quad \int_{\Omega_{e_{i}}}\boldsymbol{U}_{hp}\cdot \boldsymbol{w}^{hp}_{x}dx\approx  \int_{\Omega_{e_{i}}}\boldsymbol{U}\cdot \boldsymbol{w}^{hp}_{x}dx.\]

Combining the above approximations and setting $\mathscr{X} = \sum_{\Omega_{e_{i}}\in\mathscr{T}_{h}}\mathscr{X}_{i}$, while defining the inner product \[(\boldsymbol{a}_{hp}^{n},\boldsymbol{b}_{hp})_{\Omega_{\mathcal{G}}} = \sum_{\Omega_{e_{i}}\in\mathscr{T}_{hp}}\int_{\Omega_{e_{i}}}\boldsymbol{a}_{hp}^{n}\cdot\boldsymbol{b}_{hp} dx,\] we arrive at our approximate solution to (\ref{system2}) as the pair of functions $(\boldsymbol{U}_{hp},\boldsymbol{\Sigma}_{hp})$ for all $t\in (0,T)$ satisfying: \begin{center}\underline{The Discontinuous Galerkin formulation}\end{center}
\begin{equation}
\begin{aligned}
\label{aprox}
& a) \ \boldsymbol{U}_{hp}\in C^{1}([0,T); S_{h}^{p}), \ \ \boldsymbol{\Sigma}_{hp}\in S_{h}^{p}, \\
& b) \ \frac{d}{dt}(\boldsymbol{U}_{hp},\boldsymbol{v}_{hp})_{\Omega_{\mathcal{G}}}+\tilde{\Phi}(\boldsymbol{U}_{hp},\boldsymbol{v}_{hp}) -  \Theta(\boldsymbol{U}_{hp},\boldsymbol{v}_{hp}) \\ & \qquad\qquad-\mathscr{G}(\boldsymbol{\Sigma}_{hp},\boldsymbol{U}_{hp},\boldsymbol{v}_{hp})+ \mathscr{N}(\boldsymbol{\Sigma}_{hp},\boldsymbol{U}_{hp},\boldsymbol{v}_{hp})=0, \\  & c) \ \mathscr{Q}(\hat{\boldsymbol{U}},\boldsymbol{\Sigma}_{hp},\boldsymbol{U}_{hp},\boldsymbol{w}_{hp},\boldsymbol{w}_{x}^{hp}) = 0, \\
& d) \ \boldsymbol{U}_{hp}(0)=\Pi_{hp}\boldsymbol{U}_{0},
\end{aligned}
\end{equation} where $\Pi_{hp}$ is a projection operator onto the space of discontinuous piecewise polynomials $S_{h}^{p}$.  Below we utilize the standard $L^{2}$--projection, given for a function $\boldsymbol{f}_{0}\in L^{2}(\Omega_{e_{i}})$ such that our approximate projection $\boldsymbol{f}_{0,h}\in L^{2}(\Omega_{e_{i}})$ is obtained by solving, $\int_{\Omega_{e_{i}}}\boldsymbol{f}_{0,h}\boldsymbol{v}_{hp} dx = \int_{\Omega_{e_{i}}}\boldsymbol{f}_{0}\boldsymbol{v}_{hp} dx$.

The discretization in time follows now directly from (\ref{aprox}), where we employ a family of SSP (strong stability preserving, or often ``total variation diminishing (TVD)'') Runge-Kutta schemes as discussed in \cite{Ruuth,SO}.  That is, for the generalized SSP  Runge-Kutta scheme we rewrite (\ref{aprox}$b$) in the form: $\mathbf{M}\boldsymbol{U}_{t} = \mathbf{R}$, where $\boldsymbol{U} = (\boldsymbol{U}_{1},\ldots,\boldsymbol{U}_{p})$ for each element from (\ref{shapefunctions}), where $\mathbf{R}=\mathbf{R}(\boldsymbol{U},\boldsymbol{\Sigma})$ is the  advection-diffusion contribution along with the source term, and where $\mathbf{M}$ is the usual mass matrix.  Then the generalized $s$ stage of order $\gamma$ SSP Runge-Kutta method (denoted SSP($s,\gamma$) or RKSSP($s,\gamma$)) may be written to satisfy: \begin{equation}\begin{aligned}\label{SSPRK} & \boldsymbol{U}^{(0)}  = \boldsymbol{U}^{n}, \\ & \boldsymbol{U}^{(i)} = \sum_{r = 0}^{i-1}\left(\alpha_{ir}\boldsymbol{U}^{r}+\Delta t\beta_{ir}\mathbf{M}^{-1}\mathbf{R}^{r}\right), \quad\mathrm{for} \ \ i=1,\ldots,s \\ & \boldsymbol{U}^{n+1} =\boldsymbol{U}^{(s)},\end{aligned}\end{equation} where $\mathbf{R}^{r} = \mathbf{R}(\boldsymbol{U}^{r},\boldsymbol{\Sigma}^{r})=\mathbf{R}\left(\boldsymbol{U}^{r},\boldsymbol{\Sigma}^{r},t^{n}+\delta_{r}\Delta t \right)$ and the solution at the $n$--th timestep is given as $\boldsymbol{U}^{n}=\boldsymbol{U}_{|t=t^{n}}$ and at the $n$--th plus first timestep by $\boldsymbol{U}^{n+1}=\boldsymbol{U}_{|t=t^{n+1}}$, with $t^{n+1}=t^{n}+\Delta t$.  The $\alpha_{ir}$ and $\beta_{ir}$ are the coefficients arising from the Butcher Tableau, and the third argument in $\mathbf{R}^{r}$ corresponds to the time-lag complication arising in the constraints of the TVD formalism.  That is $\delta_{r}=\sum_{l=0}^{r-1}\mu_{rl}$, where $\mu_{ir} = \beta_{ir}+\sum_{l=r+1}^{i-1}\mu_{lr}\alpha_{il}$, where we have taken that $\alpha_{ir}\geq 0$ satisfying $\sum_{r=0}^{i-1}\alpha_{ir}=1$.

Our examples in this paper will all be given in the context of the \emph{discontinuous Galerkin} shallow water code described in \cite{KubatkoWD,KBDWM,BKWD,KDW,KWD,Dawson2010}, which employs a fully coupled system of (\ref{aprox}) including coupled eddy viscosity, time varying free boundary data, coupled chemical reactor models, \emph{etc.} to the shallow water system of equations.  For the polynomial basis we choose the hierarchical Dubiner basis, and our meshes are comprised of triangular elements.

\section{\texorpdfstring{\protect\centering $\S 3$ Types of $p$-enrichment}{\S 3  Types of $p$-enrichment}}

Here we present a family of generalizable dynamic $p$-enrichment schemes based on both local and global data.  Our first class of methods effectively extend the formalisms presented in \cite{Mich2} to what we refer to here as \emph{fixed tolerance schemes}.  These schemes use only local data to estimate the regularity of the solution, and then adapts the solution based on a fixed scalar--valued global tolerance setting.  Next we introduce the class of \emph{dioristic schemes}.  In these schemes, the solution is adapted based on global bounds on the variation which are chosen with respect to properties of the local regularity.  Here, the smoothness estimators from \cite{Mich2} are extended to treat the solution vector $\boldsymbol{U}_{hp}$ as a conjunction of weakly coupled components which each have their own variational bounds with respect to a set global tolerance.  This variation in smoothness is then used to determine the stabilization regime.

\subsection{\texorpdfstring{$3.1$ The fixed tolerance approach}{$3.1$ The fixed tolerance approach}}

Let us first present the class of dynamic--in--$p$ enrichment schemes that rely on the local solution data and are characterized by a global fixed tolerance.  Here we may state  our general goal as being one of two things: (1) to locate within the solution areas of ``excess'' variation and to probe these areas in order to capture higher order structure, or (2) to identify areas of potentially destabilizing variation in the solution, and to delimit these areas for de-enrichment in keeping with the prescribed stability conditions of the solution (\emph{e.g.} the CFL condition).  We will discuss these more below.

The first type of enrichment scheme, which we refer to here simply as the \emph{Type I fixed tolerance scheme}, apply to solutions in which substantial regularity (\emph{e.g.}  $\boldsymbol{U}\in C^{\infty}$) might be assumed over the entire domain $\Omega\times[0,T]$.  In these situations, the regularity of the solution is to be exploited such that we aim to resolve the ``areas of highest interest,'' which are those areas which show maximal variation of the solution $\boldsymbol{U}_{hp}$.  The Type I scheme is particularly attractive in application models where local high energy behavior is the signature of an ``area of interest.''  For example, in coastal models of hurricane storm surge, where one might want to resolve wave models along bay and river inlets in order to recover high accurate flooding behavior, this type of scheme might be particularly well-suited, as narrow channels often tend to focus the relative energy signature of a local subregion. 

To be more precise, the Type I scheme takes the approximate solution vector $\boldsymbol{U}_{hp}$ and computes an auxiliary sensor $\Pi$ (which we may also think of as a ``relative smoothness'' or ``relative regularity'' sensor) over each $i$-th component of $\boldsymbol{U}_{hp}$ (where $i = 1,\ldots, m$), defined by:  \begin{equation}\label{smooth}\Pi^{i}_{j}=\bigg|\frac{\boldsymbol{U}_{hp}^{i}|_{\omega_{j}} - \boldsymbol{U}_{hp}^{i}|_{c}}{\chi_{j}}\bigg|,\end{equation} where the solution  $\boldsymbol{U}_{hp}$ is evaluated at $c$, the centroid of element $\Omega_{e}$, and $\omega_{j}$, the midpoint of the $j$--th face of $\Omega_{e}$. The function $\chi_{j}$ may be set to either the distance $\chi_{j} = |\omega_{j} - c|$ as in \cite{KubBDW}, or the product $\chi_{j} = \omega_{j}c$ as in \cite{BurbeauS}.  In either case, over each timestep $n$ with respect to (\ref{SSPRK}) the following $p$-enrichment functional $\mathfrak{E}_{\mathrm{I}_{e}}= \mathfrak{E}_{\mathrm{I}_{e}}(\mathscr{P}^{k}(\Omega_{e}^{n}))$ is evaluated over each cell $\Omega_{e}$: \begin{center}\setlength{\fboxsep}{15pt}\setlength{\shadowsize}{2pt}\doublebox{\begin{minipage}{6in}\begin{center}\large{\underline{\bf Type I fixed tolerance scheme}}\end{center} \begin{equation}\label{disc}\mathfrak{E}_{\mathrm{I}_{e}} = \left\{\begin{matrix}  \mathscr{P}^{k+1}(\Omega_{e}^{n}) & \mathrm{if} \ \left((\sup_{i}\sup_{j}\Pi^{i}_{j} \geq \epsilon) \land (k+1\leq p_{\max})\right) \land (\tau_{0}\geq t^{w}), \\  \mathscr{P}^{k-1}(\Omega_{e}^{n}) & \mathrm{if} \ (\inf_{i}\sup_{j}\Pi^{i}_{j} < \epsilon) \land (k-1 \geq p_{\min})\land (\tau_{0}\geq t^{w}),  \\  \mathscr{P}^{k}(\Omega_{e}^{n}) & \ \mathrm{otherwise,} \end{matrix}\right.\end{equation}\end{minipage}}\end{center} where $\tau_{0}$ is a counter that restricts the enriching/de-enriching so that it only occurs every $t^{w}\in\mathbb{N}$ timesteps, and where $k\in\{1,\ldots,p\}$.

Depending on the choice of the global tolerance $\epsilon$ --- which is just some positive number $\epsilon\in\mathbb{R}^{+}$ --- the \emph{Type I fixed tolerance scheme} (\ref{disc}) can provide either a fairly stringent or a fairly loose cutoff with respect to which elements it flags for $p$-enrichment.  The most immediate difficulty that the scheme seems to present, is how exactly to choose the value of $\epsilon$.  If too small, then the entire solution immediately climbs to $p_{\max}$ and the dynamic nature of the algorithm is lost on the domain, and nothing particularly useful is accomplished.  If the value is too large, then the solution gets trapped at $p_{\min}$ and never enriches appreciably beyond the initial $p$ state of the solution.  In fact, this issue becomes somewhat of a complication for solutions which demonstrate substantial variation over time with respect to their ``regularity profile.''  In this case, the question arises: is it better to tune the value of $\epsilon$ to the initial state of the solution?, or is it better to tune $\epsilon$ to the maximal state of the solution?, or is it better to tune $\epsilon$ to the average state of the solution over $(0,T)$?, ... and so forth.   

It should not come as a surprise here that the answers to many of these questions are quite application dependent.  However, the questions themselves reveal perhaps the most important motivation for looking into $p$-enrichment regimes which do not harbor this type of global restriction with respect to the fixed tolerance scalar $\epsilon$.  Admittedly, one could also work to generate an application dependent form of $\epsilon$ which varies appropriately with respect to the solution, as $\epsilon=\epsilon(t,\boldsymbol{x})$, in order to stay ``centered'' with respect to the average value of $\Pi^{i}_{j}$, for example.  And while such an approach may actually be quite effective for a specific model, it would also be quite difficult to generalize without addressing in some detail the form of the model-dependent regularity estimator itself.  We will revisit this issue some in \textsection{3.2}.  

The second type of enrichment scheme we would like to consider in this section is the \emph{Type II fixed tolerance scheme}.   Here, in the Type II regime, we omit any assumptions on the regularity of our solution \emph{a priori}.   Rather, in the Type II regime we assume the possibility of numerical shock fronts arising due to instabilities that might be caused by pathological perturbations in highly variable solutions, natural discontinuities developing in the nonlinear systems, artifacts arising from numerical instabilities, or the like.  

One way of formalizing when it is appropriate to choose such a regime, is to say that Type II schemes are reserved for solutions demonstrating both appreciable local gradients $\nabla_{x}\boldsymbol{U}_{hp}\neq 0$, and nonzero local mean curvature of the solution, related by $\sim\nabla_{x}\cdot\nabla_{x}\boldsymbol{U}_{hp}$ (see \cite{ST}).  As such they are designed with an eye towards solutions which have both a large dynamic range in the magnitude of the unknowns, as well as solutions with large local spatial variations.  

In order to fully develop the Type II schemes, we again develop a regularity functional for $\boldsymbol{U}_{hp}$, but in this context we employ a slightly different formalism from the Type I schemes.  That is, the \emph{Type I fixed tolerance schemes} abstractly try to capture the amount a solution varies with respect to the element's center and the members of its boundary.  But now, we wish to try and develop an auxiliary functional $\Pi^{i}_{j}$ which takes into account the order of the variation over the element.

Perhaps the most obvious way of developing such a functional would be to utilize the jump condition between the base element and its neighbors.  After all, if the solution varies dramatically between two cells in a discontinuous basis with respect to the jump condition, then clearly the order of the variation is high in that area.  For example, in \cite{BurbeauS} the Van Leer minmod function is utilized to define the auxiliary sensor \begin{equation}\label{minmod}\Pi^{i}_{j}=\mathrm{minmod}(\boldsymbol{U}^{i}_{hp}|_{v_{j}^{+}}-\boldsymbol{U}^{i}_{hp}|_{c},\boldsymbol{U}^{i}_{hp}|_{v_{j}^{-}}-\boldsymbol{U}^{i}_{hp}|_{c}), \end{equation} where $v_{j}$ is the $j$--th vertex of $\Omega_{hp}$ evaluated with respect to the standard jump condition.  As $\Pi_{i}^{j}\to 0$ the solution becomes smoother, and one may subsequently employ an algorithm similar to (\ref{disc}) (up to the direction of inequalities, for example).  

This method offers a nice solution to the problem, but it introduces two factors which we wish to avoid in our present context.  The first is, it introduces a nonlocal stencil so that the auxiliary functional (\ref{minmod}) does not strictly depend on only the information contained within a local element.  Which leads to the second issue, being: the auxiliary functional (\ref{minmod}) would then be most effective when the mesh itself is aligned along the lines of maximal variation.  But this is precisely what occurs in many standard $h$-refinement algorithms (\emph{e.g.} see \cite{MES,Dem}), which use the solution behavior across element boundaries in order to refine the mesh along potential discontinuities --- a feature considered a requirement in order to obtain the requisite exponential convergence of $hp$-adaptive regimes.  

However, in the context of an $hp$-adaptive regime trying to $h$-refine and $p$-enrich a cell simultaneously can rapidly lead to unstable behavior.  This is impossible to avoid when using the same regularity indicator for both $h$ and $p$, unless one adopts one of two alternative approaches: either $h$-refine the cell while $p$-de-enriching it --- leading to an automatic counteraction in the accuracy of the local solution, which is unnecessary --- or, choose two different tolerances $\epsilon$ for the $h$-refinement and $p$-enrichment, respectively.  While this may work in practice, it erroneously seems to imply that the role of $h$ and $p$ are exactly the same, but just ``act'' at different magnitudes.  Since this is not the case (particularly at discontinuities), we look to develop a measure which isolates the unique properties of $p$-enrichment more explicitly.

Towards this goal, we define a local (cell-dependent) regularity estimator that relies on the weighted norm of the higher order components of the solution.  That is, let $\breve{\boldsymbol{U}}_{hp}$ be the evaluated solution in the basis where the highest components have been truncated.  In other words the members of $\mathscr{P}^{k-1}(\Omega_{e}^{\gamma})$ in our hierarchical basis.  Then we take the norm of the higher order components (\emph{i.e.} all those coefficients corresponding to $k>k-1$), and quotient out by the norm of the solution at $k$, defining the local regularity estimator: \begin{equation}\label{smoothdisc}\Pi_{i}^{e} = \left(\frac{\|\boldsymbol{U}^{i}_{hp}-\breve{\boldsymbol{U}}^{i}_{hp}\|_{L^{q}(\Omega_{e})}}{\|\boldsymbol{U}^{i}_{hp}\|_{L^{q}(\Omega_{e})}}\right),\quad\mathrm{for}\quad \breve{\boldsymbol{U}}^{i}_{hp}\in\mathscr{P}^{k-1}(\Omega_{e}^{\gamma})\quad\mathrm{and} \quad \boldsymbol{U}^{i}_{hp}\in\mathscr{P}^{k}(\Omega_{e}^{\gamma}).  \end{equation}  Here we have taken the standard Lebesgue $L^{q}$ norms (except when $q=2$ in which case we preferentially take the standard inner product). 

Now our regularity estimator (\ref{smoothdisc}) can be viewed as a weighted average of the nonlinear coefficients of the basis (at least whenever $p_{\min} = 1$).
We use this function then to develop our Type II algorithm, which is designed to truncate higher order oscillations by way of $p$-de-enrichment, while $p$-enriching areas which show high regularity.  That is, in the \emph{Type II fixed tolerance scheme}, we replace (\ref{disc}) with the Type II functional:  \begin{center}\setlength{\fboxsep}{15pt}\setlength{\shadowsize}{2pt}\doublebox{\begin{minipage}{6in}\begin{center}\large{\underline{\bf Type II fixed tolerance scheme}}\end{center} \begin{equation}\label{disc2}\mathfrak{E}_{\mathrm{II}_{e}} = \left\{\begin{matrix}  \mathscr{P}^{k+1}(\Omega_{e}^{n}) & \mathrm{if} \ (\sup_{i}\log_{10}\Pi^{e}_{i}\leq A) \land (k+1\leq p_{\max})\land (\tau_{0}\geq t^{w}), \\  \mathscr{P}^{k-1}(\Omega_{e}^{n}) & \mathrm{if} \ (\inf_{i}\log_{10}\Pi^{e}_{i} \geq  A) \land (k-1 \geq p_{\min}),  \\  \mathscr{P}^{k}(\Omega_{e}^{n}) & \ \mathrm{otherwise,} \end{matrix}\right.\end{equation}\end{minipage}}\end{center} where the bound satisfies \begin{equation} A = \left\{ \begin{matrix} \log_{10}\tilde{c}k^{-q^{2}}+c, & \mathrm{for} \ p> p_{\min} \\ \sup_{i}\log_{10}\Pi^{e}_{i}, & \mathrm{otherwise} \end{matrix}\right.\end{equation} such that $\tilde{c},c\in\mathbb{R}^{+}$ are user defined constants, where $c\in(0,10)$ is recommended (see for example \cite{WanMa}) for resolving discontinuities in the context of $hp$-adaptivity, and where we have found $\tilde{c}\in (-2,2)$ optimal.  The basic intuition that underpins the use of (\ref{smoothdisc}) is the observation that the coefficients in the basis are assumed to decay at a rate comparable to that of the Fourier coefficients in a standard expansion of the solution --- which clearly decay at a rate of  $1/k^{4}$ for $q=2$ (see \cite{WanMa,PMH,PPe}), to which we obtain an indicator of the relative local regularity of the solution, \emph{i.e.} the faster the coefficients decay, the more regular the local solution.  Thus we obtain equation (\ref{smoothdisc}), which approaches zero as the solution becomes smoother, and where setting $\tilde{c}>0$ is a sharper restriction than the more permissive (though substantially less stable) $\tilde{c}\leq 0$.

The \emph{Type II fixed tolerance scheme} can be very stabilizing due to the fact that given an appropriate choice of $A$ it will always de-enrich with respect to the elements experiencing the maximal variation.  However, as with the  \emph{Type I fixed tolerance scheme} (\ref{disc}), finding the correct value for $A$  can become a very subtle procedure, and the choice may ultimately be far from ideal over the full solution domain $\Omega\times [0,T]$.  Moreover, the Type II regime may not satisfy the objective of the enrichment scheme in the context of the given application, where one might be more concerned with fleshing out structure in areas of higher variation, rather than optimizing the algorithm to enrich most stably.

Nevertheless, the existence of solutions which may vary substantially over either space or time (or both) underscores the need to develop schemes which do not rely on globally fixed space and time independent constants.  Though a number of approaches have been developed along these line, here we present a novel set of algorithms which inherently rely upon a ``stabilizing center'' condition, and to which we refer to conventionally as: dioristic enrichment schemes.

\subsection{\texorpdfstring{$3.2$ The dioristic algorithms}{$3.2$ The dioristic algorithms}}

The class of dioristic enrichment schemes do not rely on knowing any global properties of the solution ahead of time (\emph{e.g.} knowing the global bounds on the local variation of the maximal and minimal components in order to choose an appropriate tolerance such as $\epsilon$ or $A$, as required in \textsection{3.1}), but is designed to maximize the available computational resources in such a way as to distribute \emph{as many enriched cells} as possible with respect to a parsing of the total local variation present within the components of the state vector around a ``stabilizing center.''  In order to achieve this, we must define what is meant by a ``stabilizing center,'' as well as develop a functional representation of our solution which couples each component of the state variable in a way that consistently identifies the behavior of the local variation.

A particularly nice way of developing such a fully coupled functional over $\boldsymbol{U}_{hp}$ is to derive a unique energy functional $\mathscr{S}=\mathscr{S}(t,\boldsymbol{x})$ for the system.  However, deriving such a functional $\mathscr{S}$ requires both choosing the exact form of the system \emph{a priori} (often including, for example, fully explicit forms for the boundary conditions) as well as knowing quite a lot about the form of that equation (\emph{e.g.} analytic existence and regularity of solutions).  Thus, these types of methods are far more difficult to generalize.  As an alternative we wish to work in a more general framework, where we assume that the formal entropy of the system may not be easily available or implementable \emph{a priori}, and so we resort to a slightly more weakly coupled form of the energy functionals.  Let us refer the reader to \cite{MES} for more details on how to construct entropy($\mathscr{S}$)-stabilized $hp$-adaptive regimes by way of formal energy methods.

We now proceed by defining what is meant in the present context by a ``stabilizing center.''  Since our goal is to develop an efficient and easily generalizable $p$-enrichment methodology, we assume from here forward that we may only use the information at timestep $t^{n}$ from either $\boldsymbol{U}_{hp}$ or the derived regularity estimators $\Pi(\boldsymbol{U}_{hp})$ in order to determine our dioristic functionals.  To do this we first define the range $\xi=\xi(\Pi(\boldsymbol{U}_{hp}))$ of the variation in the regularity of each component $i$ of the solution vector $\boldsymbol{U}_{hp}$ as $\xi_{i} = (\max_{\Omega_{\mathcal{G}}}\Pi_{i} - \min_{\Omega_{\mathcal{G}}}\Pi_{i})$.  That is, by the global maximum $\max_{\Omega_{\mathcal{G}}}(\cdot)$ we simply mean $\max_{\Omega_{\mathcal{G}}}=\max_{\forall\Omega_{e}\in\Omega_{\mathcal{G}}}(\cdot)$, and so forth.  Next we set the composite function $\delta_{i}=\delta_{i}(\xi_{i},\mu_{i})$ of the range $\xi_{i}$ and the adjustable weight $\mu_{i}\in (0,1)$, such that the product $\delta_{i}=\mu_{i}\xi_{i}$ determines a weighted distance of the range.   

Then denoting the global average smoothness of in each component $i$ of the solution vector as $\mathrm{Avg}_{\Omega_{\mathcal{G}}}\Pi_{i}$, we define the ``stabilizing center'' at timestep $n$ as the discrete subdomain $\mathfrak{c}\subseteq\Omega_{hp}$ comprised of the union of elements over which the solution satisfies the condition, \begin{equation}\label{stabcen} \mathfrak{c} = \Bigg\{ \bigcup_{1\leq j\leq ne}\Omega_{e_{j}} :  \big|\Pi_{i} - \mathrm{Avg}_{\Omega_{\mathcal{G}}}\Pi_{i}\big| < \delta_{i}, \ \forall i\Bigg\}.\end{equation}  The subdomain $\mathfrak{c}$ is ``stable'' in the sense that the mean value of the regularity of the solution must be commensurate to the stability settings of the free parameters associated to the formulation of (\ref{aprox}) and (\ref{SSPRK}).  Informally what this means is that ``on average'' the free parameter settings (\emph{e.g.} $\Delta t$, $dx$, \emph{etc.}) must be chosen such that the stability conditions (\emph{e.g.} the CFL condition, \emph{etc.}) are ``on average'' satisfied.  In fact, this should be viewed as an implicit condition on both of the dioristic algorithms, which can be stately more heuristically as simply assuming that the stabilizing center $\mathfrak{c}$ is in fact stable.

Then we are able to determine the Type I dioristic functional $\mathfrak{D}_{\mathrm{I}_{e}}^{\delta_{i}}$ as complementary to (\ref{disc}).  That is here, as in \textsection{3.1}, the Type I functional aims to enrich areas of greater variability using $\Pi^{i}_{j}$ from (\ref{smooth}).  But in the dioristic setting, the enriching about a fixed global parameter $\epsilon$ is replaced by an enriching about the ``stabilizing center'' $\mathfrak{c}$.  That is, over each component $i$ of the state vector $\boldsymbol{U}_{hp}$ the Type I dioristic functional $\mathfrak{D}_{\mathrm{I}_{e}}^{\delta_{i}}$ is set to satisfy:  \begin{center}\setlength{\fboxsep}{15pt}\setlength{\shadowsize}{2pt}\doublebox{\begin{minipage}{6.2in}\begin{center}\underline{\bf Type I dioristic scheme}\end{center}\begin{equation}\label{discd}\mathfrak{D}_{\mathrm{I}_{e}}^{\delta_{i}} = \left\{\begin{matrix}  \mathscr{P}^{k+1}(\Omega_{e}^{\gamma}) & \mathrm{if} \ \left(\big|\sup_{j}\Pi^{i}_{j} - \mathrm{Avg}_{\Omega_{\mathcal{G}}}\sup_{j}\Pi^{i}_{j} \big|\geq \delta_{i}) \land (k+1\leq p_{\max})\right) \land \left(\tau_{0}\geq t^{w}\right), \\  \mathscr{P}^{k-1}(\Omega_{e}^{\gamma}) & \mathrm{if} \ \left( \big|\sup_{j}\Pi^{i}_{j} - \mathrm{Avg}_{\Omega_{\mathcal{G}}}\sup_{j}\Pi^{i}_{j} \big| < \delta_{i} \quad \forall i\right) \land (k-1 \geq p_{\min})\land (\tau_{0}\geq t^{w}),  \\  \mathscr{P}^{k}(\Omega_{e}^{\gamma}) & \ \mathrm{otherwise,} \end{matrix}\right.\end{equation}\end{minipage}}\end{center} where we are using the same definitions as in \textsection{3.1}, except that here $\delta_{i}$ has the following functional dependencies: $\delta_{i} = \delta_{i}(\sup_{j}\Pi^{i}_{j},\mu_{i})$.  More precisely, in this context we simply define $\delta_{i} = \mu_{i}(\max_{\Omega_{\mathcal{G}}}\sup_{j}\Pi^{i}_{j} - \min_{\Omega_{\mathcal{G}}}\sup_{j}\Pi^{i}_{j})$ for each component $i$ of the state vector, where the global average smoothness takes the form, $\mathrm{Avg}_{\Omega_{\mathcal{G}}}\sup_{j}\Pi^{i}_{j}$.

The \emph{Type I dioristic scheme} may be described as an algorithm that takes the value of each component of the solution at timestep $n$, finds how close to the average value of the component over the whole domain the evaluated solution is; and then, when any component of the solution's variability exceeds that of $\mathfrak{c}$ on the element, it locally $p$-enriches the solution.  On the other hand, the Type I dioristic functional $\mathfrak{D}_{\mathrm{I}_{e}}^{\delta_{i}}$ only de-enriches the solution if the element lies within $\mathfrak{c}$, thus having the effect of reducing the average polynomial degree in $\mathfrak{c}$.  This feature is developed to attempt to counterbalance the destabilizing effect that $p$-enriching the solution only in the areas of highest variability can have over even a relatively uniform space of solutions.   

For the Type II dioristic functional $\mathfrak{D}_{\mathrm{II}_{e}}^{\delta_{i}}$ we develop a similar approach contextualized with respect to the ``stabilizing center'' $\mathfrak{c}$.  The aim of the \emph{Type II dioristic scheme} is to develop a scheme which takes the elements within the stabilizing center $\mathfrak{c}$ and elevates their polynomial order, while flagging those elements in the extremal regions (\emph{i.e.} $\Omega_{e_{i}}\notin \mathfrak{c}$) of $\Omega_{hp}$ for de-enrichment.

Here again we use the composite function $\delta_{i}$ for each component $i$ of the state vector $\boldsymbol{U}_{hp}$, but now we define the global average regularity of each component of the solution by, $\mathrm{Avg}_{\Omega_{\mathcal{G}}}\log_{10}\Pi_{i}^{e}$.  Then similar to the Type I dioristic functional, we define:  \begin{center}\setlength{\fboxsep}{15pt}\setlength{\shadowsize}{2pt}\doublebox{\begin{minipage}{6in}\begin{center}\underline{\bf Type II dioristic scheme}\end{center} \begin{equation}\label{discd2}\mathfrak{D}_{\mathrm{II}_{e}}^{\delta_{i}} = \left\{\begin{matrix}  \mathscr{P}^{k+1}(\Omega_{e}^{r}) & \mathrm{if} \ \left(\big|\log_{10}\Pi^{e}_{i}- \mathrm{Avg}_{\Omega_{\mathcal{G}}}\log_{10}\Pi_{i}^{e}\big| < \delta_{i} \quad \forall i\right) \land (k+1\leq p_{\max})\land (\tau_{0}\geq t^{w}), \\  \mathscr{P}^{k-1}(\Omega_{e}^{r}) & \mathrm{if} \  \left(\big|\log_{10}\Pi^{e}_{i}- \mathrm{Avg}_{\Omega_{\mathcal{G}}}\log_{10}\Pi_{i}^{e}\big| \geq  \delta_{i}\right) \land (k-1 \geq p_{\min}),  \\  \mathscr{P}^{k}(\Omega_{e}^{r}) & \ \mathrm{otherwise,} \end{matrix}\right.\end{equation}\end{minipage}}\end{center} where now we have functional dependencies given by $\delta_{i}=\delta_{i}(\Pi^{e}_{i},\mu_{i})$, and defined by $\delta_{i}=  \mu_{i}(\max_{\Omega_{\mathcal{G}}}\log_{10}\Pi^{e}_{i} - \min_{\Omega_{\mathcal{G}}}\log_{10}\Pi^{e}_{i})$.

The Type II algorithm obeys a certain type of parsimony with respect to its enriching functionality, where the only way for an element to get enriched is if the solution vector is smooth enough on the element to become a member of stabilizing center $\mathfrak{c}$.  Then, within these islands of stability the polynomial order is increased.  Everywhere else in the domain, the solution is de-enriched, operating under the tacit assumption that elements $\Omega_{e_{i}}\notin\mathfrak{c}$ possess either potentially destabilizing irregularity, or, relative to the rest of the solution, are ``nearly constant'' so that the higher order structural information recovered represents a meager benefit to the global behavior of the solution.

Notice that both the Type I and Type II dioristic algorithms require setting the local parameter $\mu_{i}$.  Since the parameter $\delta_{i}$ in both (\ref{discd}) and (\ref{discd2}) is determined with respect to the range $\xi_{i}$, the parameter $\mu_{i}$ just represents the weighted distance with respect to the total variation of the regularity at timestep $n$ that the cell based regularity may vary from the global average regularity in order to remain a member of the stabilizing center.  It is also possible to allow $\mu_{i}$ to depend on time $\mu_{i}=\mu_{i}(t)$, though due to the strong time--dependence of $\xi_{i}=\xi_{i}(t)$ this is usually unnecessary in practice, and, as the parameters in \textsection{3.1}, the behavior of $\mu_{i}$ in time is quite subtle to try to accurately predict. 

Nevertheless, both dioristic algorithms use the componentwise local variation in the solution space with respect to the the global variation in that component at a particular time $t^{n}$ to determine which areas in the domain are --- relatively speaking --- experiencing the largest relative fluctuations.  Thus for well-behaved solutions the dioristic functionals $\mathfrak{D}_{\mathrm{I}_{e}}^{\delta_{i}}$ and  $\mathfrak{D}_{\mathrm{II}_{e}}^{\delta_{i}}$  can be interpreted as energy-type functionals which sense the local variation in each component of the solution, and then $p$-enrich the domain with respect to some weighted percentage of the relative variation in each component as determined by a choice of $\mu_{i}$, while trying to stabilize by simultaneously de-enriching other components of the solution with respect to a ``stabilizing center.''

\section{\texorpdfstring{\protect\centering $\S 4$ Some numerical examples}{\S 4 Some numerical examples}}

We now consider a number of examples in order to test and analyze the behaviors of the various $p$-enrichment schemes presented in \textsection{3}.  The first example is designed as a difficult problem with many symmetries, where the variation of the solution is quite large, existing at the precipice of numerical shock formation.  The second example is quite complicated and is more of an application model, which attempts to realistically model estuary eutrophication in the Gulf of Mexico.

\subsection{\texorpdfstring{$4.1$ Contaminant plume declinature}{$4.1$ Contaminant plume declinature}}

\begin{figure}[t!]
\centering
\includegraphics[width=11.0cm]{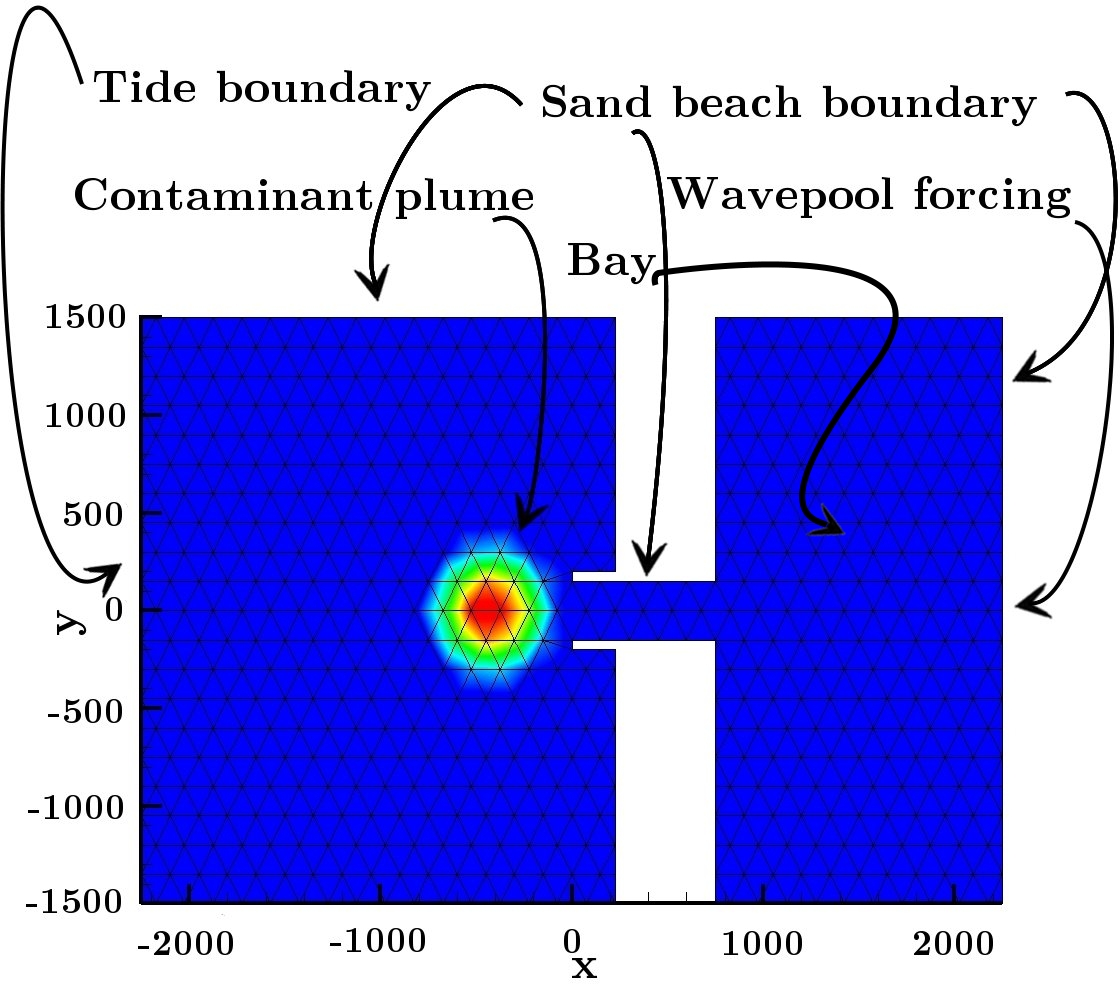} \\ 
\caption{Initial conditions, boundary conditions and course mesh of the contaminant plume declinature model, where the bay is on the right with a \emph{wavepool forcing}.}
\label{fig:plume}
\end{figure}

Our first example is that of a plume declinature model.  The underlying physics that the model tries to address --- with respect to the fairly idealized setting in two dimensions --- is, consider a bay with a contaminant plume on the far side of an inlet into the bay (see Figure \ref{fig:plume}).  Then we ask the physical question: is it possible to prevent the contaminant plume from entering the bay by imparting mechanical wave energy on the near side of the the bay using a standard hydraulic wavepool generator?  As we show below in two dimensions and given a very powerful local hydraulic generator at least, the answer is yes.  We additionally address the issue of $p$ accuracy in the $p$-adapting regime, versus the standard convergence-in-$p$ of the solution in a slightly more restrained setting.

\begin{table}[h]
\centering 
\begin{tabular}{|c | c | c | c | c | c | c | c | c | c | c | c |}
\hline
$p$ &  Type &  $L^{1}$-error ($\iota_{hp}$)  & $t^{w}$ & $\epsilon$ & $\tilde{c}$ & $c$ & $q$ & $\mu$ & $\Delta t$ (s) & SSP & rCPU \rule{0pt}{3ex} \rule[0ex]{0pt}{0pt} \\ 
\hline\hline
7 & --  & -- & -- & -- & -- & -- & -- & -- & $2.50\times 10^{-1}$ & (6,4) & 1.00 \rule{0pt}{3ex} \rule[0ex]{0pt}{0pt}  \\
\hline
5 & --  & $1.01\times 10^{-4}$ & -- & -- & -- & -- & -- & -- & $5.00\times 10^{-1}$ & (6,4) & 0.29 \rule{0pt}{3ex} \rule[0ex]{0pt}{0pt}  \\
\hline
4 & --  & $1.56\times 10^{-4}$ & -- & -- & -- & -- & -- & -- & $5.00\times 10^{-1}$ & (6,4) & 0.21 \rule{0pt}{3ex} \rule[0ex]{0pt}{0pt}  \\
\hline
3 & --  &  $1.87\times 10^{-4}$   & -- & -- & -- & -- & -- & -- & $7.50\times 10^{-1}$ & (6,4) & 0.12 \rule{0pt}{3ex} \rule[0ex]{0pt}{0pt}  \\
\hline
2 & --  &  $2.82\times 10^{-4}$ & -- & -- & -- & -- & -- & -- & $7.50\times 10^{-1}$ & (6,4) & 0.11 \rule{0pt}{3ex} \rule[0ex]{0pt}{0pt}  \\
\hline
1 & --  &  $3.42\times 10^{-4}$ & -- & -- & -- & -- & -- & -- & $1.00$ & (6,4) & 0.07 \rule{0pt}{3ex} \rule[0ex]{0pt}{0pt}  \\
\hline
1--2 & $\mathfrak{E}_{\mathrm{I}_{e}}$  & $3.37\times 10^{-4}$ & 10 & $5\times 10^{-5}$ & -- & -- & -- & -- & $7.50\times 10^{-1}$ & (3,2) & 0.08 \rule{0pt}{3ex} \rule[0ex]{0pt}{0pt}  \\
\hline
2--3 & $\mathfrak{E}_{\mathrm{I}_{e}}$  & $1.88\times 10^{-4}$ & 10 & $5\times 10^{-6}$ & -- & -- & -- & -- & $7.50\times 10^{-1}$ & (3,2) & 0.09 \rule{0pt}{3ex} \rule[0ex]{0pt}{0pt}  \\
\hline
1--2 & $\mathfrak{D}^{\delta_{i}}_{\mathrm{I}_{e}}$  & $2.98\times 10^{-4}$ & 10 & -- & -- & -- & -- & $\frac{1}{100}$ & $7.50\times 10^{-1}$ & (3,2) & 0.10 \rule{0pt}{3ex} \rule[0ex]{0pt}{0pt}  \\
\hline
2--3 & $\mathfrak{D}^{\delta_{i}}_{\mathrm{I}_{e}}$  & $1.88\times 10^{-4}$ & 10 & -- & -- & -- & -- & $\frac{1}{100}$ & $7.50\times 10^{-1}$ & (3,2) & 0.11 \rule{0pt}{3ex} \rule[0ex]{0pt}{0pt}  \\
\hline
1--3 & $\mathfrak{E}_{\mathrm{II}_{e}}$  & $3.37\times 10^{-4}$ & 10  & --  & 1  & $\frac{1}{2}$ & 2 & -- & 2.00 & (3,2) & 0.03 \rule{0pt}{3ex} \rule[0ex]{0pt}{0pt}  \\
\hline
2--4 &  $\mathfrak{E}_{\mathrm{II}_{e}}$   &  $2.56\times 10^{-4}$ & 10 & -- & 1 & $\frac{1}{2}$ & 2 & -- & 1.50 & (3,2) & 0.06 \rule{0pt}{3ex} \rule[0ex]{0pt}{0pt}  \\
\hline
1--2 & $\mathfrak{D}^{\delta_{i}}_{\mathrm{II}_{e}}$  & $3.36\times 10^{-4}$ & 10 & -- & -- & -- & 2 & $\frac{21}{100}$ & $2.00$ & (3,2) & 0.04 \rule{0pt}{3ex} \rule[0ex]{0pt}{0pt}  \\
\hline
2--3 & $\mathfrak{D}^{\delta_{i}}_{\mathrm{II}_{e}}$  & $2.75\times 10^{-4}$ & 10 & -- & -- & -- & 2 & $\frac{21}{100}$ & $1.5$ & (3,2) & 0.06 \rule{0pt}{3ex} \rule[0ex]{0pt}{0pt}  \\
\hline

\end{tabular}
\caption{We give the normalized $L^{1}$-error of $\iota_{hp}$ after half a day (\emph{i.e.} $T= 0.5$ days) of simulation time with respect to the $p=7$ converging in $p$ solution using $a_{force}=0$ m, with the relative CPU times (rCPU) of the approximate solutions  over continuously adapted (\emph{e.g.} $p=1$-$3$ implies that $p$ spans $\{1,2,3\}$) solutions coupled to the the enrichment schemes. }
\label{table:declineacc}
\end{table}

For the solution to the contaminant declinature problem, consider the transport form of the two-dimensional shallow water initial--boundary problem, determined by: \begin{equation}\begin{aligned}\label{shallwater}&\partial_{t}(\zeta\iota) + \nabla_{x}(H\iota\boldsymbol{u}) = 0, \\ \partial_{t}(H\boldsymbol{u}) + \nabla_{x}&\mathfrak{S} + S - \eta\Delta_{x}(H\boldsymbol{u})- g\zeta\nabla_{x}h = 0, \\  \mathfrak{S} = & \left(H\boldsymbol{u}\otimes\boldsymbol{u} + \frac{1}{2}g(H^{2}-h^{2})\right),\end{aligned}\end{equation} where the initial data satisfies \[\zeta_{t=0}=\zeta_{0},\quad\boldsymbol{u}_{t=0}=\boldsymbol{u}_{0},\quad \iota_{|t=0}=\iota_{0}.\]  

Here the solution space is comprised of the velocity field $\boldsymbol{u}=\boldsymbol{u}(t,\boldsymbol{x})$, the elevation of the free surface $\zeta=\zeta(t,\boldsymbol{x})$, the total water column height $H=\zeta+b$, where $b=b(\boldsymbol{x})$ is the time independent bathymetric depth (measured positive downwards), and $\iota=\iota(t,\boldsymbol{x})$ represents the relative concentration of a chemically inert contaminant.  The terms containing the gravitational constant $g\in\mathbb{R}^{+}$ correspond to those deriving from the frequently employed hydrostatic pressure assumption, while the constant $\eta\in\mathbb{R}^{+}$ is the eddy viscosity coefficient (here set to $\eta =10 \ \mathrm{m}^{2}\cdot\mathrm{s}^{-1}$), and $S$ corresponds to the remaining source terms, which generally could contain Coriolis forces, bottom friction, wind forcing, \emph{etc}.   Here we simply set a bottom friction approximation using the Chezy formula such that $S=.0025|\boldsymbol{u}|/H$ (see \textsection{4.2} for more complicated source settings).

\begin{figure}[t!]
\centering
\includegraphics[width=15cm]{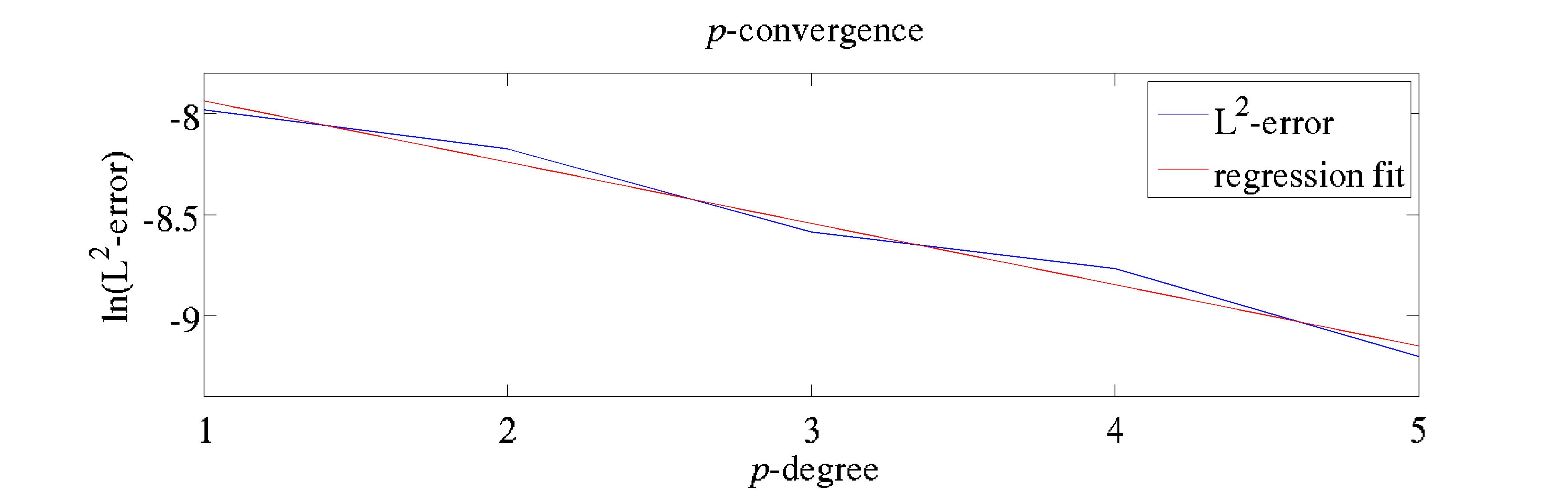}
\caption{Here we show the $p$-convergence of the solution denoted in Table \ref{table:declineacc}, where the \emph{static forcing} is used.}
\label{pconv}  
\end{figure}

The initial data is $\zeta_{0}=0$ m, $b=10$ m, and $\boldsymbol{u}_{0}=0$ m$\cdot$s$^{-1}$ with initial relative contaminant concentration: \[\iota_{0} =  1/5, \ \ \mathrm{for} \ 0 \leq r < a_{1}, \quad \mathrm{where}\quad r = \sqrt{(x+a)^{2} + y ^{2}}, \ a=420 \ \mathrm{m}, \ \mathrm{and} \ a_{1} = 250 \ \mathrm{m}.\]  It is not difficult to see that the system (\ref{shallwater}) is easily formulated in the form of (\ref{system2}), where here we use a local Lax–-Friedrichs flux, and where we use an upwinding scheme for the coupled contaminant transport model.

\begin{table}[t!]
\centering 
\begin{tabular}{|c | c | c | c | c | c | c | c | c | c | c | c |}
\hline
$p$ &  Type &  $L^{1}$-error ($\iota_{hp}$)  & $t^{w}$ & $\epsilon$ & $\tilde{c}$ & $c$ & $q$ & $\mu$ & $\Delta t$ (s) & SSP & rCPU \rule{0pt}{3ex} \rule[0ex]{0pt}{0pt} \\ 
\hline\hline
7 & --  & -- & -- & -- & -- & -- & -- & -- & $2.50\times 10^{-1}$ & (6,4) & 1.00 \rule{0pt}{3ex} \rule[0ex]{0pt}{0pt}  \\
\hline
1--7 & $\mathfrak{E}_{\mathrm{II}_{e}}$  & $8.42\times 10^{-6}$ & 0  & --  & 1  & 1 & 2 & -- & 2.00 & (3,2) & 0.07 \rule{0pt}{3ex} \rule[0ex]{0pt}{0pt}  \\
\hline
1--7 & $\mathfrak{E}_{\mathrm{II}_{e}}$  & $8.47\times 10^{-6}$ & 0  & --  & 1  & 1 & 3 & -- & 1.50 & (3,2) & 0.09 \rule{0pt}{3ex} \rule[0ex]{0pt}{0pt}  \\
\hline
1--7 & $\mathfrak{E}_{\mathrm{II}_{e}}$  & $8.52\times 10^{-6}$ & 10  & --  & 1  & 1 & 2 & -- & 2.00 & (3,2) & 0.06 \rule{0pt}{3ex} \rule[0ex]{0pt}{0pt}  \\
\hline
1--7 & $\mathfrak{E}_{\mathrm{II}_{e}}$  & $8.51\times 10^{-6}$ & 20  & --  & 1  & 1 & 2 & -- & 2.00 & (3,2) & 0.06 \rule{0pt}{3ex} \rule[0ex]{0pt}{0pt}  \\
\hline
1--7 & $\mathfrak{E}_{\mathrm{II}_{e}}$  & $8.52\times 10^{-6}$ & 40  & --  & 1  & 1 & 2 & -- & 2.00 & (3,2) & 0.06 \rule{0pt}{3ex} \rule[0ex]{0pt}{0pt}  \\
\hline
1--7 &  $\mathfrak{E}_{\mathrm{II}_{e}}$   &  $8.49\times 10^{-6}$ & 0 & -- & 1 & -1 & 2 & -- & $5.00\times 10^{-1}$ & (5,3) & 0.80 \rule{0pt}{3ex} \rule[0ex]{0pt}{0pt}  \\
\hline
1--7 &  $\mathfrak{E}_{\mathrm{II}_{e}}$   & $8.22\times 10^{-6}$ & 20 & -- & 1 & -1 & 2 & -- & $5.00\times 10^{-1}$ & (5,3) & 0.44 \rule{0pt}{3ex} \rule[0ex]{0pt}{0pt}  \\
\hline
5--7 & $\mathfrak{E}_{\mathrm{II}_{e}}$ & $3.50\times 10^{-6}$ & 0 & -- & 1 & -1 & 2 & -- & $2.50\times 10^{-1}$ & (5,3) & 1.74 \rule{0pt}{3ex} \rule[0ex]{0pt}{0pt}  \\
\hline
1--7 & $\mathfrak{D}^{\delta_{i}}_{\mathrm{II}_{e}}$  & $8.52\times 10^{-6}$ & 0 & -- & -- & -- & 2 & $\frac{1}{5}$ & $5.00\times 10^{-1}$ & (3,2) & 0.26 \rule{0pt}{3ex} \rule[0ex]{0pt}{0pt}  \\
\hline
1--7 & $\mathfrak{D}^{\delta_{i}}_{\mathrm{II}_{e}}$  & $8.52\times 10^{-6}$ & 0 & -- & -- & -- & 2 & $\frac{1}{5}$ & $5.00\times 10^{-1}$ & (5,3) & 0.41 \rule{0pt}{3ex} \rule[0ex]{0pt}{0pt}  \\
\hline
1--7 & $\mathfrak{D}^{\delta_{i}}_{\mathrm{II}_{e}}$  & $8.52\times 10^{-6}$ & 10 & -- & -- & -- & 2 & $\frac{1}{5}$ &  $5.00\times 10^{-1}$ & (5,3) & 0.42 \rule{0pt}{3ex} \rule[0ex]{0pt}{0pt}  \\
\hline
1--7 & $\mathfrak{D}^{\delta_{i}}_{\mathrm{II}_{e}}$  & $8.52\times 10^{-6}$ & 20 & -- & -- & -- & 2 & $\frac{1}{5}$ & $5.00\times 10^{-1}$ & (5,3) & 0.41 \rule{0pt}{3ex} \rule[0ex]{0pt}{0pt}  \\
\hline
1--5$^{*}$ & $\mathfrak{E}_{\mathrm{I}_{e}}$  & $1.06\times 10^{-4}$ & 10 & $5\times 10^{-5}$ & -- & -- & -- & -- & $5.00\times 10^{-1}$ & (5,3) & 0.48 \rule{0pt}{3ex} \rule[0ex]{0pt}{0pt}  \\
\hline
1--5$^{*}$ & $\mathfrak{D}^{\delta_{i}}_{\mathrm{I}_{e}}$  & $1.39\times 10^{-4}$ & 10 & -- & -- & -- & -- & $\frac{1}{4}$ & $5.00\times 10^{-1}$ & (5,3) & 0.48 \rule{0pt}{3ex} \rule[0ex]{0pt}{0pt}  \\

\hline
5 & --  &  $8.49\times 10^{-6}$ & -- & -- & -- & -- & -- & -- & $5.00\times 10^{-1}$ & (5,3) & 0.40 \rule{0pt}{3ex} \rule[0ex]{0pt}{0pt}  \\
\hline
1  & --  & $8.76\times 10^{-6}$ & -- & -- & -- & -- & -- & -- & 2.00 & (3,2) & 0.02 \rule{0pt}{3ex} \rule[0ex]{0pt}{0pt}  \\
\hline

\end{tabular}
\caption{We show the robustness of the $p$-enrichment, as well as the effects on the accuracy of the \emph{weak entropy} in strong boundary forcings.  Here we show the normalized $L^{1}$-error of $\iota_{hp}$ after 2 days (\emph{i.e.} $T= 2$ days) of simulation time with respect to the $p=7$ solution using $a_{force}=6$ m, with the relative CPU times (rCPU) of the approximate solutions.  Here $^{*}$ means the solution is run using an adapting-in-$p$ slopelimiter (discussed below). }
\label{table:declinerob}
\end{table}

The boundary conditions are separated out so that we have a union of Dirichlet conditions over the total domain boundary: $\partial\Omega_{h} = \partial\Omega_{sand}\cup\partial\Omega_{tide}\cup\partial\Omega_{force}$.   Each component of $\boldsymbol{U}_{hp}$ is set independently, but due to the physical constraints on the system can be simplified to satisfy, \begin{equation}\begin{aligned} \label{bccd}& \zeta_{b}  = (\zeta_{b,1})_{sand}\cup(\zeta_{b,2})_{tide}\cup(\zeta_{b,3})_{force}, \quad \iota_{b} = (\iota_{b,123})_{sand,tide,force},\\ & \quad\qquad\boldsymbol{u}_{b} = (\boldsymbol{u}_{b,n,1}\cdot\boldsymbol{n}+\boldsymbol{u}_{b,\tau,1}\cdot\boldsymbol{\tau}) _{sand}\cup (\boldsymbol{u}_{b,n,2}\cdot\boldsymbol{n}+\boldsymbol{u}_{b,\tau,2}\cdot\boldsymbol{\tau}) _{tide} \\ & \quad\qquad\qquad\qquad\qquad\cup  (\boldsymbol{u}_{b,n,3} \cdot\boldsymbol{n}+\boldsymbol{u}_{b,\tau,3}\cdot\boldsymbol{\tau} )_{force}.\end{aligned}\end{equation}  The first boundary type $\partial\Omega_{sand}$ corresponds to a sand beach boundary condition, the second $\partial\Omega_{tide}$ to an open ocean tidal inlet condition, and the third $\partial\Omega_{force}$ to either a hydraulic wavepool periodic forcing condition, or a static condition (see below for details).  Here, as in (\ref{bounds}), $\boldsymbol{\tau}$ is the tangent unit vector and $\boldsymbol{n}$ is the outward pointing unit vector.  

Spatially, the entire west edge of the domain as labeled in Figure \ref{fig:plume} is comprised of a tidal inlet condition $\partial\Omega_{tide}$ along 20 elements, and is 3000 m long.  The center of the east edge of the bay --- which is defined as the two east edge elements with $|y|\leq 150$ m --- is given the forcing condition $\partial\Omega_{force}$.  All of the remaining 92 edges of the domain boundary $\partial\Omega_{h} $ are given the sand beach boundary condition $\partial\Omega_{sand}$.

\begin{figure}[t!]
\centering
\includegraphics[width=8.5cm]{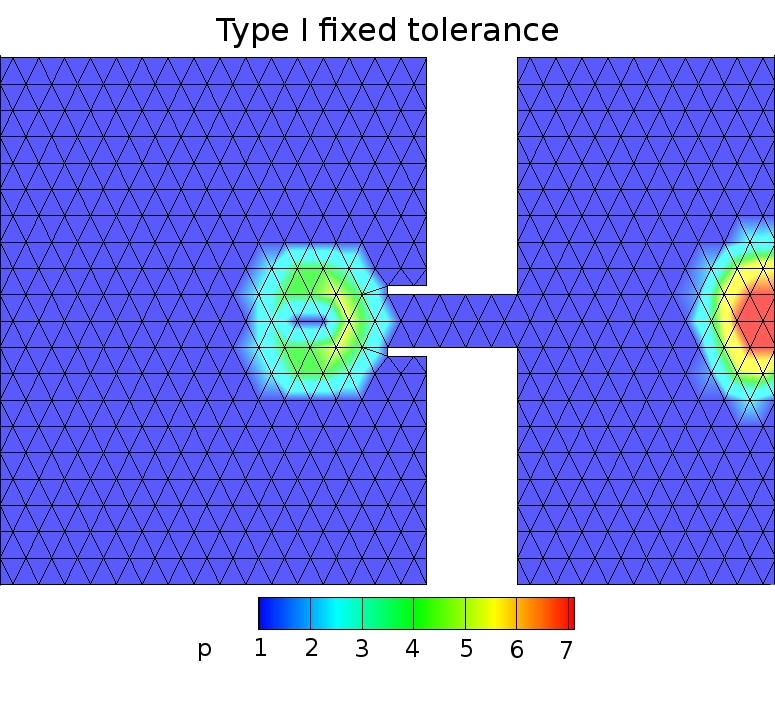} \ \includegraphics[width=8.5cm]{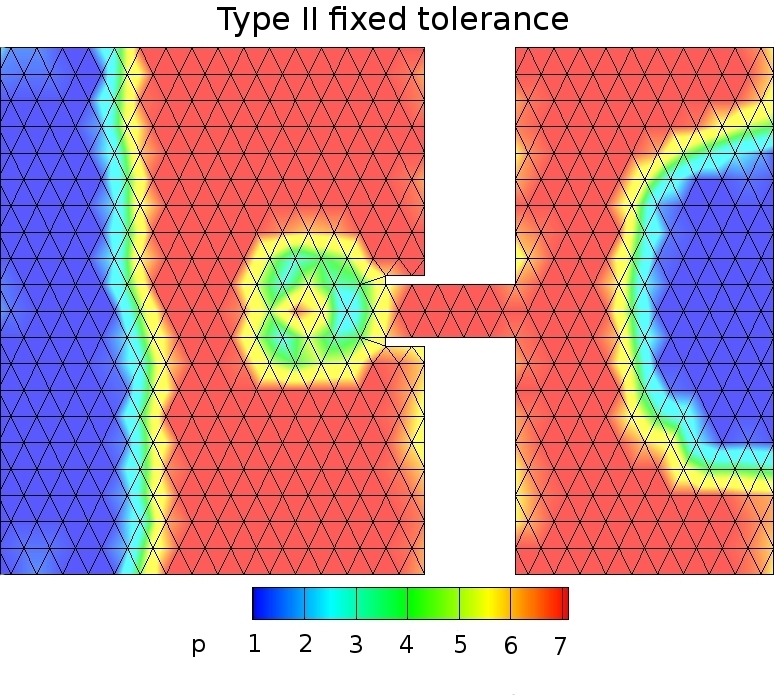}
\caption{Here we show the $p$ value of the Type I and Type II fixed tolerance algorithms after $T=42$ seconds with respect to the \emph{wavepool forcing} and the settings from Table \ref{table:p_track}.}
\label{fig:p_ad}  
\end{figure}

Let us define these three different boundary types precisely.   The first order transmissive scalar boundary values from (\ref{bccd}) are determined by the value on the elements interior at the boundary, so that $\zeta_{b,1}|_{\partial\Omega_{e_{i}}} = \zeta|_{\partial\Omega_{e_{j}}}$ and $\iota_{b,2}|_{\partial\Omega_{e_{i}}} =  \iota|_{\partial\Omega_{e_{j}}}$.  The scalar forcings at the boundary, on the other hand, are set so that for $l\in\{1,3\}$ they satisfy: \begin{equation}\begin{aligned}\label{contam}\zeta_{b,2}|_{\partial\Omega_{e_{i}}} = a_{tide}\cos (\omega_{tide} & t), \quad \zeta_{b,3}|_{\partial\Omega_{e_{i}}} = a_{force}|\cos (\omega_{force} t)|, \quad  \mathrm{and}\quad\iota_{b,l}|_{\partial\Omega_{e_{i}}} = 0.\end{aligned}\end{equation} The amplitude and frequency of the tide are given as $a_{tide}=0.75 \ \mathrm{m}$ and $\omega_{tide}= 7.02\times 10^{-5} \ \mathrm{rad}\cdot\mathrm{s}^{-1}$ respectively, while the amplitude and frequency of the \emph{wavepool forcing} is given by $a_{force}= 6 \ \mathrm{m}$ and $\omega_{force} = 2.56\times 10^{-3} \ \mathrm{rad}\cdot\mathrm{s}^{-1}$ respectively, while the \emph{static forcing} is determined uniquely by $a_{force}= 0 \ \mathrm{m}$.  Both amplitudes are ramped up over half a day in order to avoid shocking the system, where a tide of three quarters of a meter occurs every half day over $\partial\Omega_{tide}$, and a very large and localized wave of six meters is generated every twenty minutes over $\partial\Omega_{force}$ with respect to the bay (see Figure \ref{fig:plume}).   The third condition on the relative contaminant concentration in (\ref{contam}) represents first a filtration condition at the wavepool inlet, and second an absorption assumption at beach boundaries (\emph{e.g.} petroleum contaminated sand \cite{IMS}).

Finally, the velocity boundary conditions are set to obey a dampening with respect to the sandy beach and an open ocean quenching condition, while the wavepool vector is chosen as to quench the nonlinear velocities forced by $\zeta|_{\partial\Omega_{e_{i}}}$.   More clearly, for $l\in\{1,2,3\}$ we have that $\boldsymbol{u}_{b,n,l}|_{\partial\Omega_{e_{i}}}= 0 \ \mathrm{m}\cdot\mathrm{s}^{-1}$ and $\boldsymbol{u}_{b,\tau,l}|_{\partial\Omega_{e_{i}}} = 0  \ \mathrm{m}\cdot\mathrm{s}^{-1}$.  

Let us briefly discuss the results of our numerical experiments.  First, it should be clear that our boundary forcings drive our solution, and are weakly imposed as discussed above.  These \emph{weak entropy} conditions have been analyzed in previous work \cite{MESV,Martin}, and are known to be a source of error in the approximate solution.  In fact, the error on the boundary elements may scale with the mesh size $h$ in our adapting regime, such that in our case, where are boundary forcings are very strong relative to a small interior domain, the weak entropy boundary error frequently dominates.  Because of this, we do not achieve (or expect to achieve) the rates of convergence in $p$ one might expect from a manufactured solution, \emph{etc.} in the case of the \emph{wavepool forcing}, which is quite strong.  In fact, we push the boundary forcings to the edge of what the convergence-in-$p$ can even distinguish, as is clear in  Table \ref{table:declinerob} between the $p=7$, $p=5$ and $p=1$ cases; where our focus is rather the robustness of the enrichment schemes.  In Table \ref{table:declinerob} we see impressive robustness of the Type II algorithms, even with rigid constraints imposed by both the CFL condition and on the weak entropy boundary forcings, where it is clear that even running the solution in quite dynamic circumstances (\emph{e.g.} when $p\in\{1,\ldots,7\}$ and the \emph{wavepool forcing} on the boundary is applied) the solution is still stable at quite large timesteps (\emph{e..g} $\Delta t = 2$ seconds) without introducing any observable loss in accuracy.

In contrast, when we suppress the strong boundary layers given by the \emph{wavepool forcing}, and employ the \emph{static forcing} condition  $a_{force}= 0 \ \mathrm{m}$ instead, the weak entropy on the boundary is reduced and propagates below the levels of the accuracy of the interior solution.  We present these results in Table \ref{table:declineacc}.  Here we see the requisite $p$-convergence behavior of the solution with the reference solution at $p=7$ (as shown in Figure \ref{pconv}).  Notice that all four schemes under the correct settings may be used to approximate $p$-accuracy lying between integral values.  From the timestepping chosen, it is immediately clear that the Type II algorithms demonstrate greater relative stability compared to the Type I algorithms.  In fact, it should be noted that the Type I algorithms require settings that are either largely weighted towards $p=p_{\min}$ or $p=p_{\max}$.  In other words, in this test case, because the Type I algorithms enrich in the areas of greatest variation, in order to achieve stability the amount of $p$ adaptation over $T$ must be minimized.  In contrast, the Type II algorithms demonstrate substantial robustness at a number of different timesteps and settings, while still achieving $p$-convergence.  Nevertheless, it must be stressed that in some applications one strongly desires structural information in areas of maximal variation; so much so that the loss of stability in the enrichment process pays for itself in terms of recovering information with respect to the proper locale of interest.

\begin{table}[t!]
\centering
\begin{tabular}{|c | c | c | c | c | c | c | c | c | c | c | c |}
\hline
Type & $\epsilon/\mu$ & $\tilde{c}$ & $c$ & $p=1$ & $p=2$ & $p=3$ & $p=4$ & $p=5$ & $p=6$ & $p=7$ & Timestep \rule{0pt}{3ex} \rule[0ex]{0pt}{0pt} \\ 
\hline\hline
$\mathfrak{E}_{\mathrm{I}_{e}}^{*}$ & $5\times 10^{-5}$ & -- & -- & 1094 & 0 & 0 & 2 & 0 & 0 & 18 & 25 \rule{0pt}{3ex} \rule[0ex]{0pt}{0pt} \\ 
\hline
$\mathfrak{D}_{\mathrm{I}_{e}}^{\delta_{i},*}$ & $2/5$ & -- & -- & 1072 & 0 & 1 & 3 & 1 & 0 & 137  & 25 \rule{0pt}{3ex} \rule[0ex]{0pt}{0pt} \\ 
\hline
$\mathfrak{E}_{\mathrm{II}_{e}}$ & -- & 1 & $\frac{1}{2}$ & 156 & 1 & 2 & 4 & 13 & 16 & 922  & 25 \rule{0pt}{3ex} \rule[0ex]{0pt}{0pt} \\ 
\hline
$\mathfrak{D}_{\mathrm{II}_{e}}^{\delta_{i}}$ & $2/5$ & -- & -- & 124 & 20 & 7 & 4 & 9 & 9 & 941  & 25 \rule{0pt}{3ex} \rule[0ex]{0pt}{0pt} \\ 
\hline\hline
$\mathfrak{E}_{\mathrm{I}_{e}}^{*}$ & $1\times 10^{-5}$ & -- & -- & 489 & 5 & 8 & 8 & 24 & 34 & 546 & 1000 \rule{0pt}{3ex} \rule[0ex]{0pt}{0pt} \\ 
\hline
$\mathfrak{D}_{\mathrm{I}_{e}}^{\delta_{i},*}$ & $2/5$ & -- & -- & 1094 & 0 & 0 & 0 & 0 & 0 & 20  & 1000 \rule{0pt}{3ex} \rule[0ex]{0pt}{0pt} \\ 
\hline
$\mathfrak{E}_{\mathrm{II}_{e}}$ & -- & 1 & $\frac{1}{2}$ & 835 & 270 & 9 & 0 & 0 & 0 & 0  & 1000 \rule{0pt}{3ex} \rule[0ex]{0pt}{0pt} \\ 
\hline
$\mathfrak{D}_{\mathrm{II}_{e}}^{\delta_{i}}$ & $2/5$ & -- & -- & 1075 & 39 & 0 & 0 & 0 & 0 & 0  & 1000 \rule{0pt}{3ex} \rule[0ex]{0pt}{0pt} \\
\hline
\end{tabular}
\caption{Here we provide the $p$ value as computed over each element corresponding to the four $p$-enrichment algorithms from \textsection{3} given the hydraulic \emph{wavepool forcing}.  Again $^{*}$ corresponds to the use of the slopelimiter from \cite{Mich2}}
\label{table:p_track}
\end{table}

The different behavior of the fixed tolerance algorithms compared to the dioristic algorithms is also emphasized in Table \ref{table:declineacc}.  Here in the \emph{static forcing} model, our $\mathrm{M}_{2}$ tidal constituent has a period of half a day, and thus varies quite slowly with respect to the timestep (\emph{e.g.} $\Delta t = 0.5$ s implies $86400$ timesteps over $T=0.5$ days).  Nevertheless, one can see that because the energy of the solution satisfies a periodic forcing as well, the fixed tolerance regimes are only able to capture $\sim 3500$ timesteps of variation over the $86400$ timesteps.  This is because, regardless of the settings for $\epsilon$ that one uses, the ``energy'' of the solution is only within the tolerance setting for $\sim 1750$ timesteps on either side of the peak amplitude.    This behavior is demonstrated in Table \ref{table:p_track} and Figure \ref{fig:p_ad}.  The dioristic algorithms on the other hand, do a qualitatively better job of spreading out the $p$ variation over the entire time domain $[0,T)$.   Moreover, as seen in Table \ref{table:p_track}, the Type II algorithms owe a substantial component of their stability (and the robustness feature shown in Table  \ref{table:declinerob}) to the fact that the de-enrichment functional is independent of $t^{w}$, and thus is able to ``catch instabilities'' --- so to speak --- before they form; and hence they attempt to keep the average $p$ value below a ``stable setting,'' which by the CFL condition is a function of $\Delta t$. 

\begin{figure}[t!]
\centering
\includegraphics[width=8.5cm]{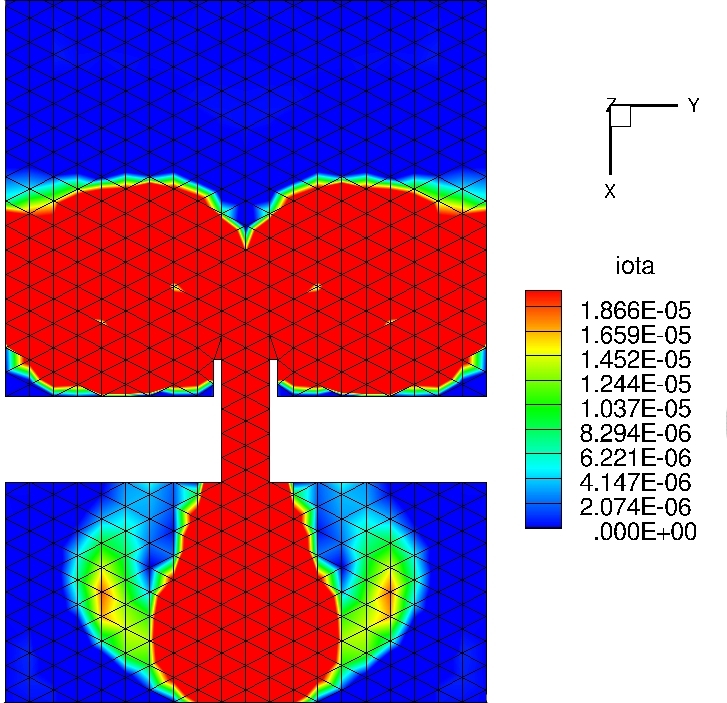} \ \includegraphics[width=8.2cm]{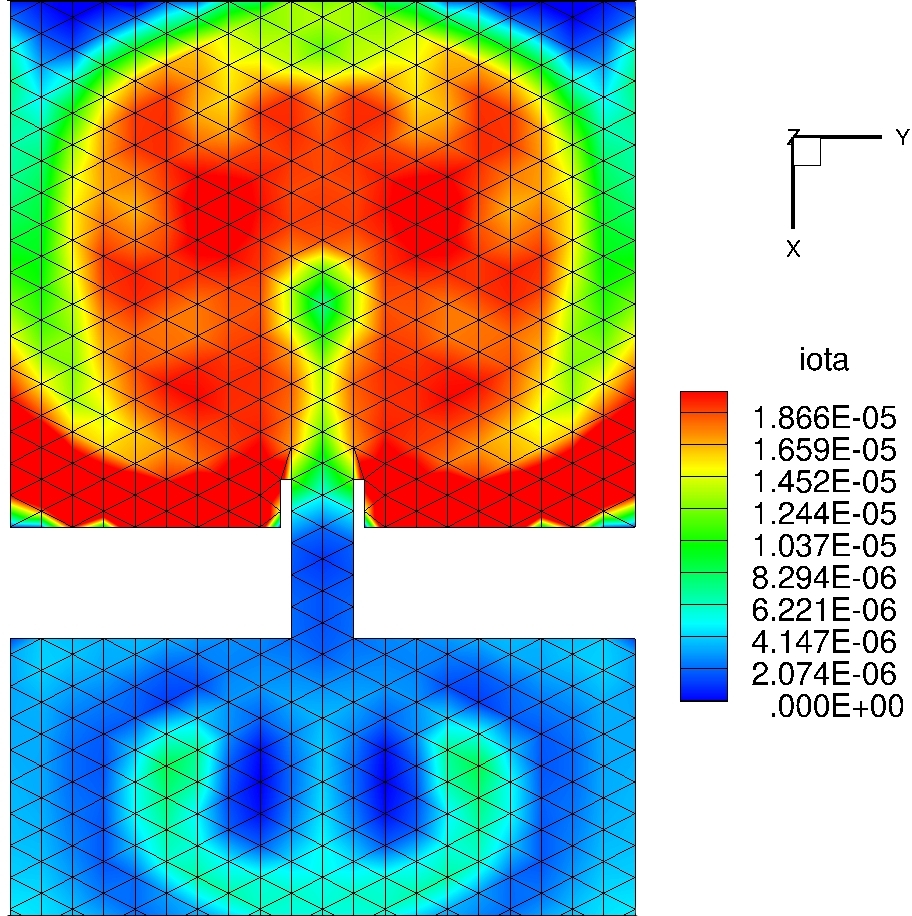}
\caption{Here we show the relative concentration of $\iota$ after $T=2$ days for the linear $p=1$ solution.  On the left is the solution with the \emph{static forcing}, and on the right is the \emph{wavepool forcing}.  It is clear that after only two days, the wavepool has rejected a large amount of the contaminant that would otherwise be in the bay.}
\label{fig:contam}  
\end{figure}

As a general trend, and as previously discussed, one might say that the Type I fixed tolerance $p$-enrichment scheme tends to $p_{\max}$ in time assuming that the energy of the system is convex.  Similarly the Type II enrichment schemes both have a tendency towards $p_{\min}$ in time under similar assumptions.  The Type I dioristic scheme, on the other hand, seems to demonstrate the most static behavior as a function of time, and it seems likely this will remain the case assuming that the signature behavior of the energy variation in $\Omega\times[0,T)$ is not significantly dampened, whether the total energy functional behavior of the system is convex or concave or some mix of the two.

It is also worth mentioning that the $p$-enrichment scheme does not come without a computational price, and in the Type I setting when $\epsilon$ is small (\emph{i.e.} sensitive) in a solution with very large gradients, the polynomial order rapidly approaches $p_{\max}$, so that when $p$ is particularly high the $p$-enrichment scheme itself also becomes more computationally expensive.  Generally we must be able to balance the computational cost of the enrichment scheme with the cost of increasing the local degrees of freedom of the solution, such that increased accuracy can be achieved without a resource cost that outweighs its worth.

Finally let us briefly discuss the physical model.  The \emph{wavepool forcing} is designed with the intent of forcing the contaminant out of the bay by way of exploiting the mechanical energy of a hydraulic wavepool generator.  Thus, we can simply test to see if, after 2 days, the amount of contaminant in the bay is reduced when employing the hydraulic wavepool generator, versus setting the static condition given by   $a_{force}= 0 \ \mathrm{m}$.  We show the results after two days in Figure \ref{fig:contam}.  First we note that the total (integrated) contaminant present in the domain in the case of the hydraulic wavepool model is $28\%$ that of the case with no wavepool generator included.  However, in the bay itself, the value is less than $18\%$ that of the case with no wavepool generator included, clearly showing substantial declinature of the contaminant in only two days.  This raises an interesting physical observation, which seems to suggest that fairly simply directed mechanical advection may offer a viable method of both declinature, and/or active transport of constituents which are inert with respect to the aqueous saline environments.  

\subsection{\texorpdfstring{$4.2$ Neutralizing estuary eutrophication}{$4.2$ Neutralizing estuary eutrophication}}

It has been generally observed that dead zones in estuarine outlets in the Gulf of Mexico correlate with algal blooms in the surrounding coastal regions.  These blooms are believed to lead to dead zones in the local marine ecology by way of large population explosions in microorganism density which substantially deplete the oxygen in the surrounding sea water, in turn leading to devastating loss of local marine life.  These blooms have been further correlated to excess nutrient levels (\emph{e.g.} inorganic esters) in the gulf watershed leading to what are known as hypereutrophic coastal regions.  In fact, the subsequent eutrophication in the local coastal region is known to be caused in part by excess relative concentrations of both phosphates and nitrates that are present largely due to the runoff pollution of man-made fertilizer in upstream agricultural centers spanning watershed riverbeds. 

\begin{figure}[t!]
\centering
\includegraphics[width=8.5cm]{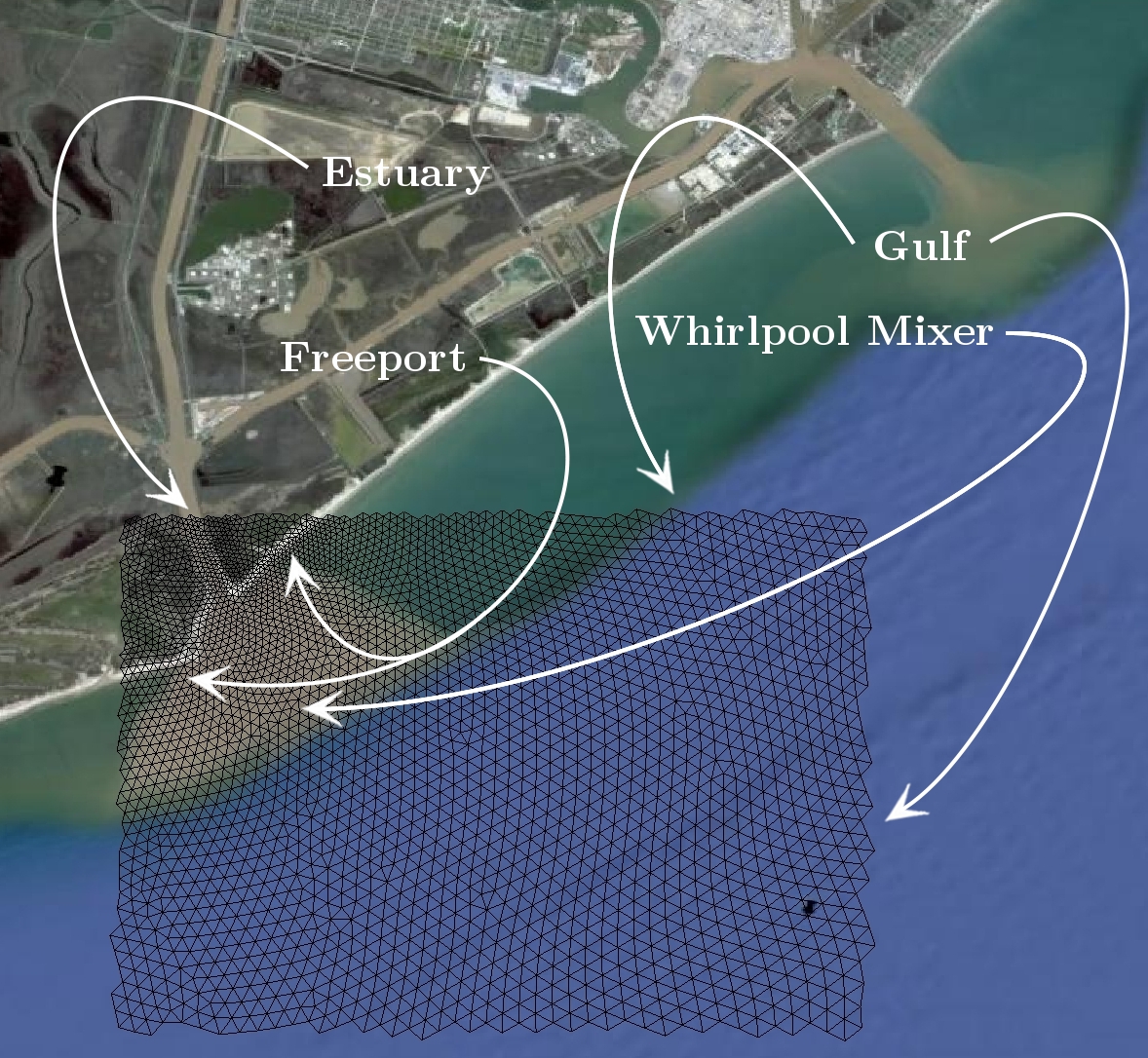} \ \includegraphics[width=8.5cm]{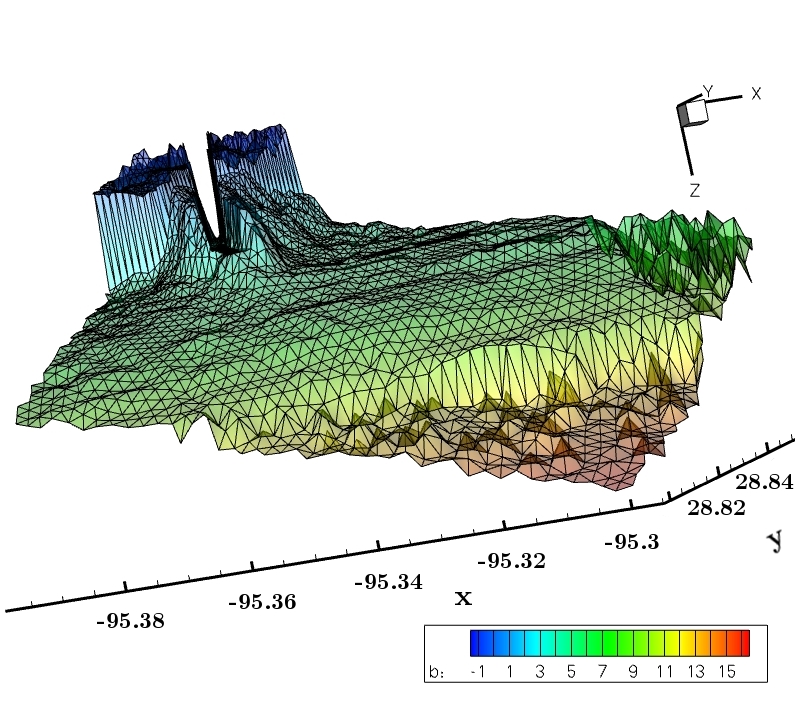}
\caption{The left shows the mesh of the Brazos estuary for the eutrophication model, overlaid on a map of a Freeport, TX satellite image compliments to Google Maps, \copyright 2009 Google, Map Data \copyright 2009 Tele Atlas.  The right gives the same mesh in terms of bathymetry, $b=b(x)$. }
\label{fig:freeport}  
\end{figure}

Here we consider a test problem designed towards developing a method of neutralizing the rate limiting constituent (\emph{i.e.} we choose phosphates as discussed in \cite{SDNMMA}, since nitrogen is often in such vast excess) at an estuarine outlet.  That is, we take the Brazos River estuary at Freeport, TX (see Figure \ref{fig:freeport}) and couple a remediation reaction to the contaminant flow from \textsection{5.2}.

Consider the chemically active form of the shallow water equations, which in conservation form satisfies: \begin{equation}\begin{aligned}\label{eutrophy}&\partial_{t}(\zeta\iota_{j}) + \nabla_{x}(H\iota_{j}\boldsymbol{u}) - H\mathscr{A}_{j} = 0, \\ \partial_{t}(H\boldsymbol{u}) + &\nabla_{x}\mathfrak{S} + S -\eta\Delta_{x}(H\boldsymbol{u})- g\zeta\nabla_{x}h = 0, \\ & \mathfrak{S} =  \left(H\boldsymbol{u}\otimes\boldsymbol{u} + \frac{1}{2}g(H^{2}-h^{2})\right), \\ \mathscr{A}_{j}=\sum_{r\in\mathfrak{R}}(&\nu_{j,r}^{b}-\nu^{f}_{j,r})\left(k_{f,r}\prod_{i=1}^{n}\iota_{i}^{\nu_{i,r}^{f}}-k_{b,r}\prod_{i=1}^{n}\iota_{i}^{\nu_{i,r}^{b}}\right)\end{aligned}\end{equation} given initial data \[\zeta_{t=0}=\zeta_{0},\quad\boldsymbol{u}_{t=0}=\boldsymbol{u}_{0},\quad \iota_{j,|t=0}=\iota_{j,0} \quad \mathrm{for} \quad j\in\{1,2\},\] where our equation is the same as that in \textsection{4.1}, except for the fact that there are now two transported reactive chemical constituents $\iota_{j}=\iota_{j}(t,\boldsymbol{x})$ for $j\in\{1,2\}$.  These constituents obey the law of mass action $\mathscr{A}_{j}=\mathscr{A}_{j}(\iota)$, while the source term $S=S(t,\boldsymbol{x})$ has been modified to include a whirlpool mixer $\mathscr{O}=\mathscr{O}(t,\boldsymbol{x})$ (as defined explicitly below) and a Coriolis parameter $\mathcal{C}=f\boldsymbol{u}H$ for a constant coriolis coefficient $f=2\Phi\sin\theta$, where $\Phi=7.292\times 10^-5 \ \mathrm{rad} \cdot \mathrm{s}^{-1}$ and $\theta=\theta(\boldsymbol{x})$ is the approximate latitude of the node.  We add these terms ($\mathscr{O}$ and $C$) to the bottom friction already present in $S$ from \textsection{4.1}.  

\begin{figure}[t!]
\centering
\includegraphics[width=12.0cm]{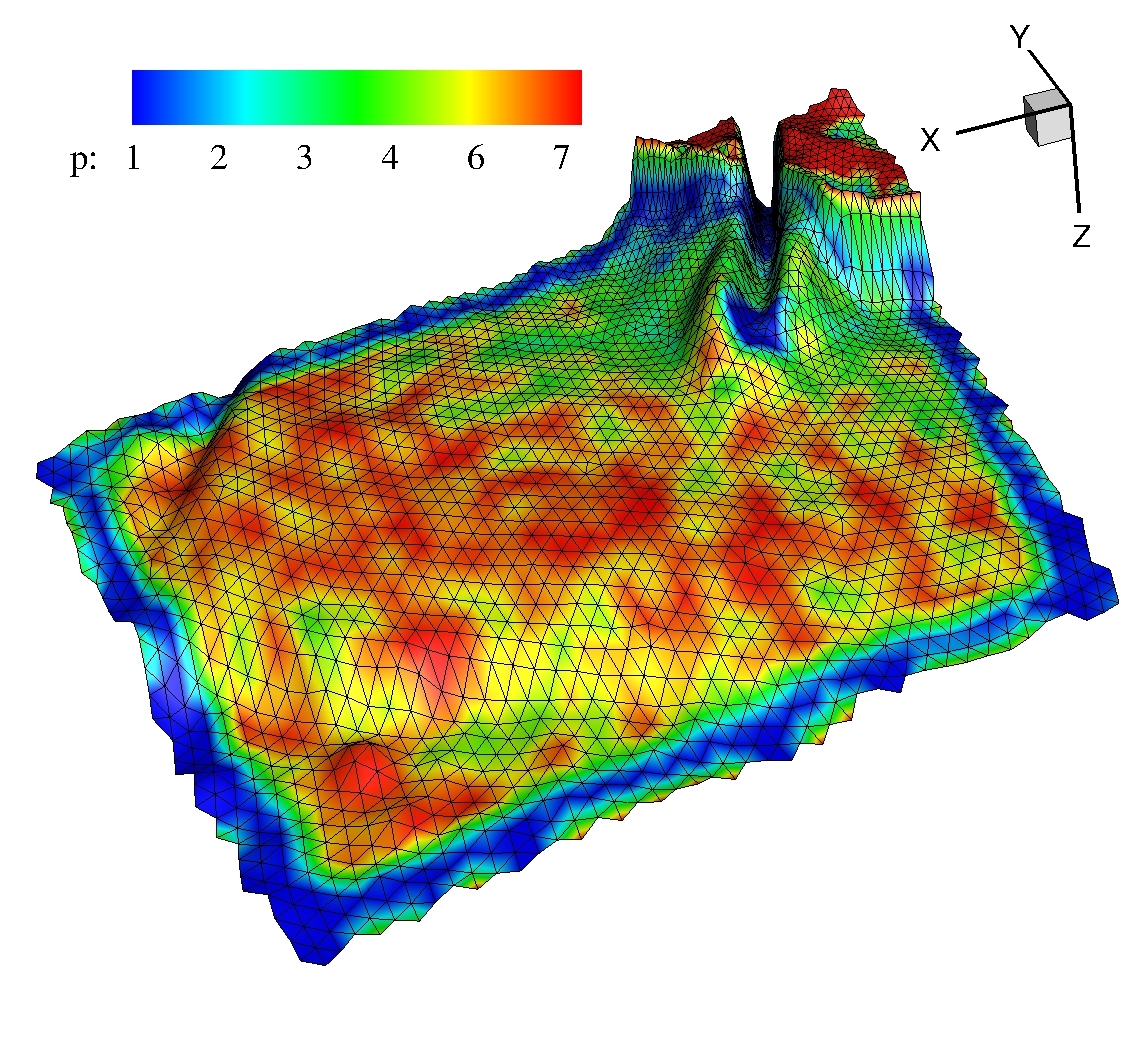}
\caption{Here we show the $p$ level of the $p=$1-7 solution to the full eutrophication problem (with whirlpool and chemistry both active) after 28 seconds, using the Type I fixed tolerance scheme, with $c=1$, $\tilde{c}=0.5$ and $\Delta t = 1$ second.  The $z$ direction is mapped with respect to the bathymetry.}
\label{fig:shortbrazos}  
\end{figure}

\begin{figure}[t!]
\centering
\includegraphics[width=8cm]{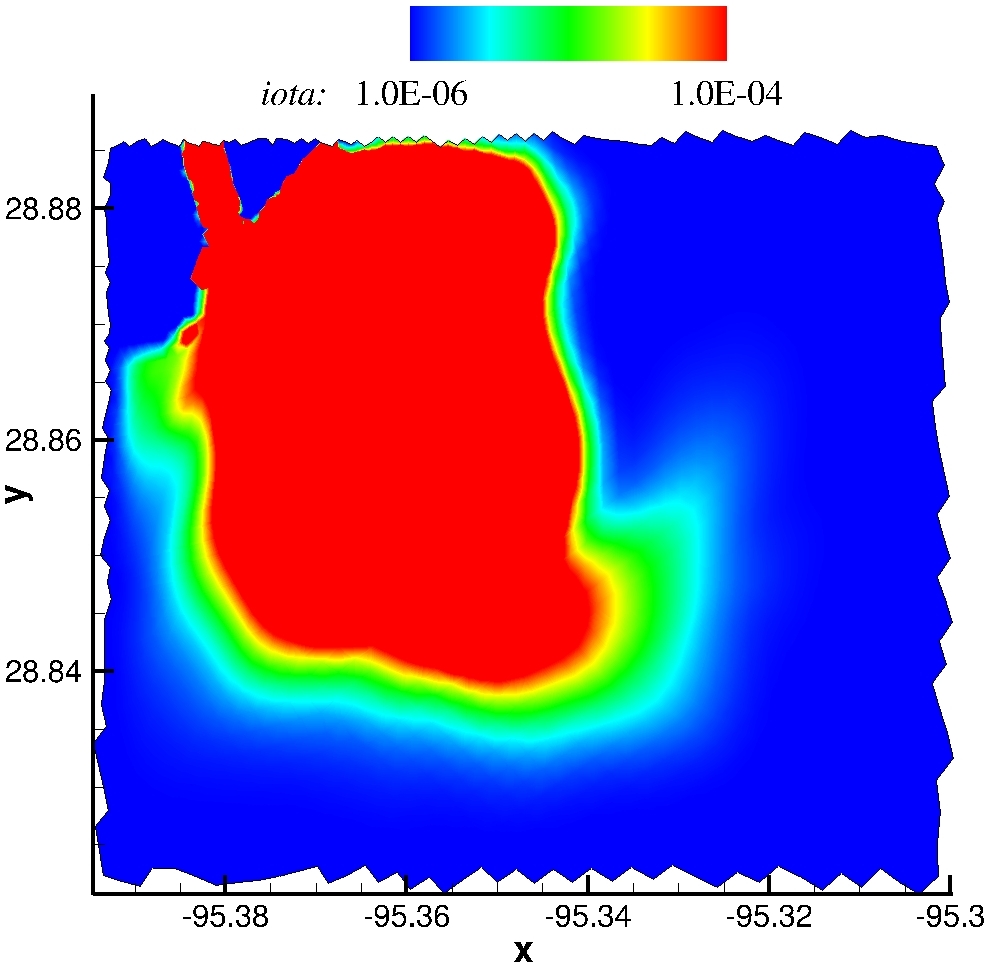} \ \includegraphics[width=8cm]{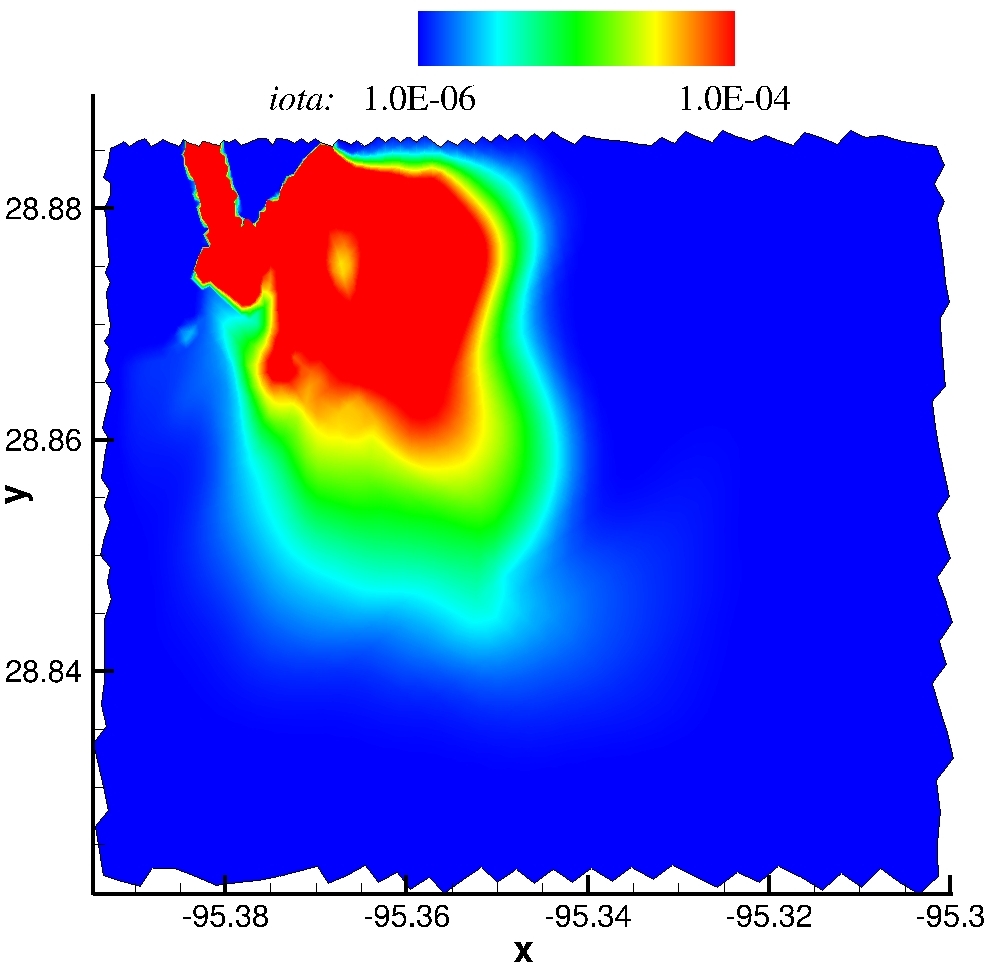} \\ \includegraphics[width=8cm]{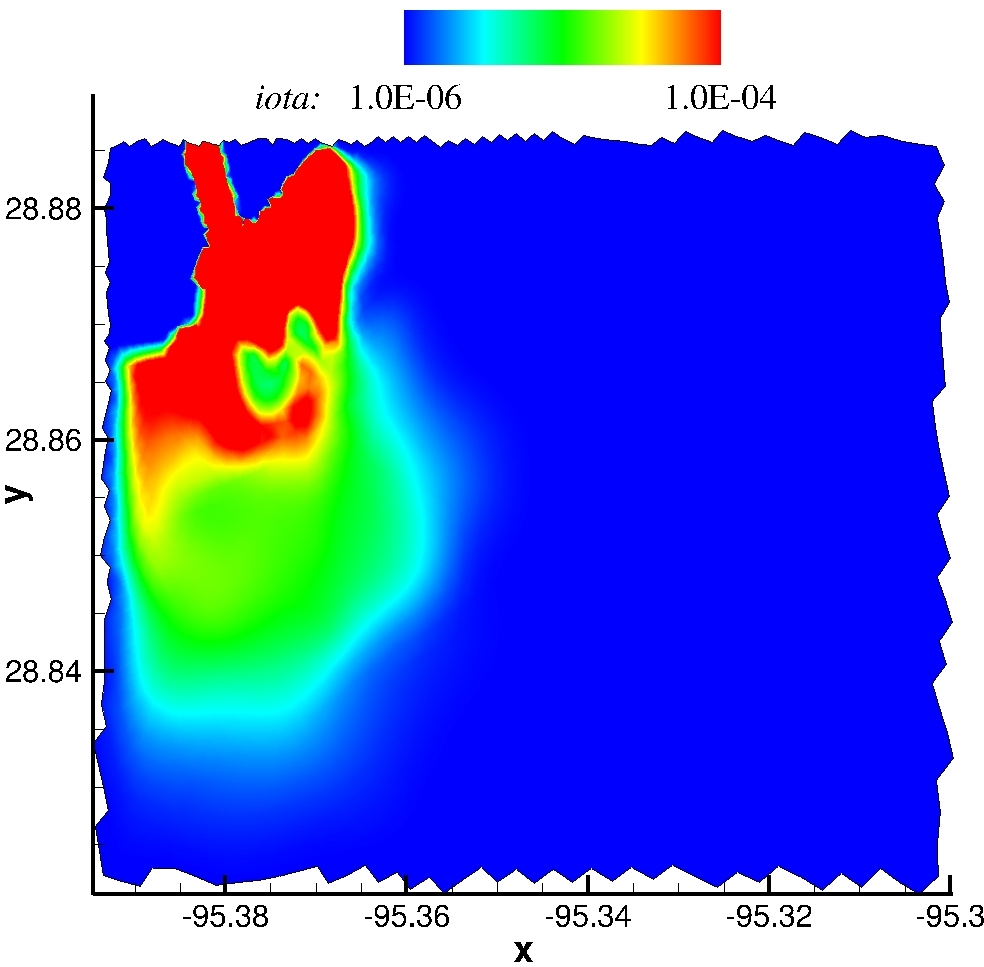} \ \includegraphics[width=8cm]{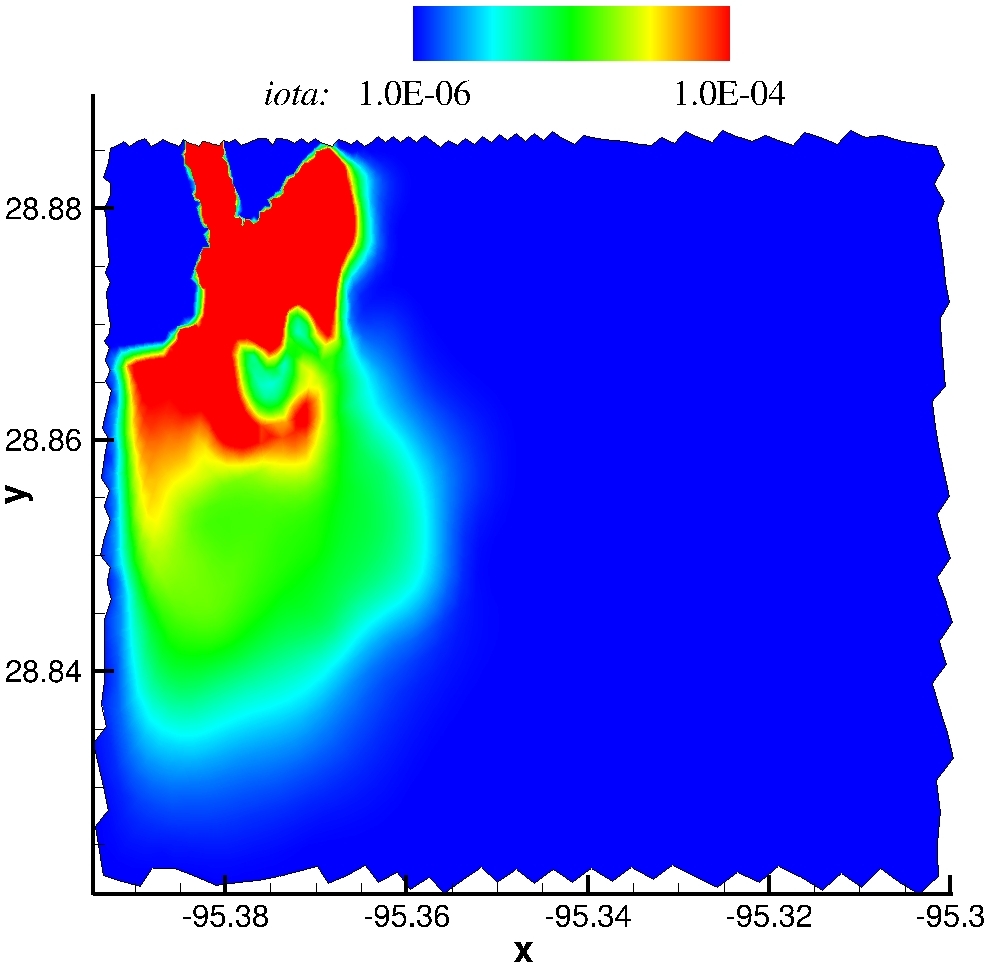}
\caption{Each shows the solution at $T=29$ days.  The upper left shows the $p=1$ solution to the inert transport problem, with chemistry and whirlpool off.  The upper right is the $p=1$ solution with chemistry on, whirlpool off.  The bottom left is the $p=1$ with chemistry on and whirlpool on.  The bottom right shows the $p\in\{1,\ldots,4\}$ Type II fixed tolerance $p$-enriched solution with $c=\tilde{c}=1$, chemistry and whirlpool turned on.}
\label{fig:freeportfull}  
\end{figure}

The chemical mass action $\mathscr{A}_{j}$ may also be viewed as a source term in the transport equation where we have neglected the usual Fickian diffusion to a first approximation, as discussed in the introduction.  Here, $k_{f,r},k_{b,r}\in\mathbb{R}^{+}$ are the forward and backward reaction rate constants, and $\nu_{j,r}^{f},\nu_{j,r}^{b}\in\mathbb{Z}$ are the corresponding constant stoichiometric coefficients given an elementary reaction $r$ in the reaction space $\mathfrak{R}$.  See \cite{MES,MS,Hirschfelder,Cheng,Giovangigli3} for more details on these basic equations.

Then, as a model problem we consider the aqueous (sea water) remediation reaction (buffered locally in the whirlpool to a $\mathrm{pH}\sim 8$) of hardened gypsum and hydrogen phosphate into mineralized hydroxyapatite and salt, as observed in \cite{Kana} by way of the proposed mechanism: \begin{equation}\label{mineraldeposits} \mathrm{CaSO}_{4}\cdot 2\mathrm{H}_{2}\mathrm{O} + \mathrm{HPO}_{4}^{2-} \ \ce{->T[{$ \ \ k_{f} \ \ $}][{NaOH}] } \ \mathrm{Na}_{2}\mathrm{SO}_{4} + \mathrm{DAp},  \end{equation} where \[  \mathrm{DAp} = \mathrm{Ca}_{10-X}(\mathrm{HPO}_{4})_{X}(\mathrm{PO}_{4})_{6-X}(\mathrm{OH})_{2-X}\cdot n\mathrm{H}_{2}\mathrm{O}\] is calcium deficient hydroxyapatite, and $X\in(0,1]$.  

Now, the gypsum $\mathrm{CaSO}_{4}\cdot 2\mathrm{H}_{2}\mathrm{O} $ is a common and cheap industrial byproduct that demonstrates good hydraulicity, while hydroxyapatite is a primary constituent of bone, is osteogenic, non-toxic and potentially easy to reclaim from the environment (if such is even desired or necessary).  The neutralization reaction of the sulfuric acid with the sodium hydroxide is taken to be diffusion limited, while the whirlpool pH is further buffered in the presence of sea water.  The reaction rate of (\ref{mineraldeposits}) is imprecise as extrapolated from \cite{KatKom,BrownP}.  As a consequence, since the estuary water temperature varies annually from 12$^{\circ}$C--30$^{\circ}$C, we take as a \emph{weak estimate} on the second order reaction rate assuming idealized conditions that $k_{f}\sim 1 \ \mathrm{L}\cdot\mathrm{mol}^{-1}\mathrm{d}^{-1}$ (though clearly this can be easily adjusted).

Being concerned only with the reactants (\ref{mineraldeposits}) may be partially decoupled from (\ref{eutrophy}) such that they satisfy the following system of ODE's: \[\begin{aligned}&\partial_{t}\iota_{1} = - k_{f}\iota_{1}\iota_{2}\quad\mathrm{and}\quad\partial_{t}\iota_{2} = - k_{f}\iota_{1}\iota_{2},\end{aligned}\] where $\iota_{1}=[\mathrm{CaSO}_{4}\cdot 2\mathrm{H}_{2}\mathrm{O}]$ and $\iota_{2}=[\mathrm{HPO}_{4}^{2-}]$ are relative concentrations.  Applying the discontinuous Galerkin reactor scheme from \cite{MES} over our discrete timestep $\Delta t = t^{n+1}-t^{n}$ we then arrive with: \[\iota_{1}^{n+1} = \iota_{1}^{n}e^{-k_{f} \iota_{2}\Delta t}\quad\mathrm{and} \quad \iota_{2}^{n+1} = \iota_{2}^{n}e^{-k_{f}\iota_{1}\Delta t},\] where it becomes easy to reclaim the relative concentration $\iota_{j}^{n+1}$ in (\ref{eutrophy}) at every timestep.  For more details on this family of DG chemical kinetic models see \cite{MES,MS}. 

The whirlpool mixer $\mathscr{O}=\mathscr{O}(t,\boldsymbol{x})$ sets the frequency of the gypsum release based on the local relative concentration of the phosphate, and also sets the speed at which the whirlpool rotates.   As such, it satisfies the following condition at timestep $n+1$:  \begin{equation}\label{while}\mathscr{O}^{n+1}= \left\{\begin{matrix}  \iota_{2,\mathscr{O}} & \mathrm{for} \ (\boldsymbol{x}_{a_{1}}<\boldsymbol{x}<\boldsymbol{x}_{b_{1}})\land (\iota_{1}\geq \iota_{1,\mathscr{O}}), \\  \boldsymbol{u}_{\mathscr{O}} & \mathrm{for} \ (\boldsymbol{x}_{a_{1}}<\boldsymbol{x}<\boldsymbol{x}_{b_{1}})\land (\boldsymbol{u}^{n}\leq \boldsymbol{u}_{\lim}),  \end{matrix}\right.\end{equation} where the whirlpool velocity is not allowed to exceed some limiting rotation velocity $\boldsymbol{u}_{\lim}$ which we set to $\boldsymbol{u}_{\lim}=0.1$ m$\cdot$s$^{-1}$, restricted to the region $(\boldsymbol{x}_{a_{1}},\boldsymbol{x}_{a_{2}},\boldsymbol{x}_{b_{1}},\boldsymbol{x}_{b_{2}})$ at the heart of the whirlpool in the mouth of the Brazos River as shown in Figure \ref{fig:freeport}.   The limiting relative concentrations are given by $\iota_{1,\mathscr{O}}=1\times 10^{-8}$ and $ \iota_{2,\mathscr{O}}=1\times 10^{-2}$, while the whirlpool velocity satisfies: \[\boldsymbol{u}_{\mathscr{O}}= r^{-1}\boldsymbol{m}(\boldsymbol{x}-\boldsymbol{x}_{h}),\quad \mathrm{for}\quad r=\sqrt{(\boldsymbol{x}-\boldsymbol{x}_{h})^{2}},\] such that the velocity increases towards the heart of the whirlpool $\boldsymbol{x}_{h}$ and the direction of the rotation is determined by the vector $\boldsymbol{m}=(m_{1},m_{2})$, setting $m_{1}=-0.05$  m$\cdot$s$^{-1}$ and $m_{2} = 0.05$ m$\cdot$s$^{-1}$.

The corresponding boundary conditions are first split into Dirichlet conditions over the domain boundary $\partial\Omega_{h} := \partial\Omega_{gulf}\cup\partial\Omega_{estuary}$ (see Figure \ref{fig:freeport}) corresponding to the Brazos River estuary and the Gulf of Mexico, except for now a free boundary subdomain is included $\partial\Omega_{land}(t,\boldsymbol{x})$ and set such that $\partial\Omega_{land}(t,\boldsymbol{x})= \partial\Omega_{land}$ corresponds to the free boundary layer at the Freeport shoreline taken with respect to a coastal wetting and drying treatment (see below).   Then by construction the total boundary $\Omega_{h,0}=\Omega_{h,0}(t,\boldsymbol{x})$ at any timestep is given as $\partial\Omega_{h,0}=\partial\Omega_{land}\cup\partial\Omega_{h}$ where the remaining boundary conditions are given by: \[\begin{aligned}& \qquad\qquad \ \zeta_{b}  = (\zeta_{b,1})_{land}\cup(\zeta_{b,2})_{gulf}\cup(\zeta_{b,3})_{estuary},  \\ & \qquad\qquad \ \ \iota_{b} =  (\iota_{b,1})_{land}\cup(\iota_{b,2})_{gulf}\cup(\iota_{b,3})_{estuary},\\ & \boldsymbol{u}_{b} = (\boldsymbol{u}_{b,n,1}\cdot\boldsymbol{n}+\boldsymbol{u}_{b,\tau,1}\cdot\boldsymbol{\tau}) _{land}\cup (\boldsymbol{u}_{b,n,2}\cdot\boldsymbol{n}+\boldsymbol{u}_{b,\tau,2}\cdot\boldsymbol{\tau}) _{gulf} \\ & \quad\qquad\qquad\qquad\cup  (\boldsymbol{u}_{b,n,3} \cdot\boldsymbol{n}+\boldsymbol{u}_{b,\tau,3}\cdot\boldsymbol{\tau} )_{estuary}.\end{aligned}\] 

First, the Brazos River estuary is treated using the following no slip hydrogen phosphate inlet conditions: \[\begin{aligned}& \zeta_{b,3}|_{\partial\Omega_{e_{i}}} = \zeta|_{\partial\Omega_{e_{j}}},\quad \boldsymbol{u}_{b,n,3}|_{\partial\Omega_{e_{i}}} = a_{estuary}\cos (\omega_{estuary}t) + 0.05 \ \mathrm{m}\cdot\mathrm{s}^{-1}, \\ & \qquad\qquad \quad   \boldsymbol{u}_{b,\tau,3}|_{\partial\Omega_{e_{i}}}= 0 \ \mathrm{m}\cdot\mathrm{s}^{-1} \quad \mathrm{and}\quad \iota_{b,3}|_{\partial\Omega_{e_{i}}} = 0.01,\end{aligned}\] where we approximate the tidal frequency by setting $\omega_{estuary}=0.729\times 10^{-4}\mathrm{rad}\cdot\mathrm{s}^{-1}$ with $\pi$ the tidal offset such that the variable amplitude per \cite{JRS} achieves a mean stream velocity of $a_{estuary} = 0.7 \ \mathrm{m}\cdot\mathrm{s}^{-1}$, and varies such that $\boldsymbol{u}_{b,n,1}|_{\partial\Omega_{e_{i}}}\in[0.02,0.12] \  \mathrm{m}\cdot\mathrm{s}^{-1}$.  Finally a phosphate inlet relative concentration of $0.01$ is taken along the estuary outflow, where here we are only interested in the relative behavior, such that up to a rescaling all concentrations are admissible and may be viewed as unitless (\emph{e.g.} volume fractions).

Next, the open ocean conditions on the Gulf of Mexico boundary are set to the following:  \[\begin{aligned}&\qquad\qquad \zeta_{b,2}|_{\partial\Omega_{e_{i}}} = \mathcal{T},\quad \boldsymbol{u}_{b,n,2}|_{\partial\Omega_{e_{i}}} =   0 \ \mathrm{m}\cdot\mathrm{s}^{-1}, \quad \boldsymbol{u}_{b,\tau,2}|_{\partial\Omega_{e_{i}}}= 0  \ \mathrm{m}\cdot\mathrm{s}^{-1}\quad\mathrm{and} \quad \iota_{b,2}|_{\partial\Omega_{e_{i}}} = 0,\end{aligned}\] where the tidal constituents function $\mathcal{T}=\mathcal{T}(t,\boldsymbol{x})$ is a linear combinations of its arguments such that $\mathcal{T} = \mathcal{T}(K_{1},O_{1},Q_{1},M_{2},S_{2},N_{2},K_{2})$, where the arguments are the usual diurnal and semidiurnal tidal constituents.   More precisely, for each constituent $\chi \in \{K_{1},O_{1},Q_{1},M_{2},S_{2},N_{2},K_{2}\}$ takes the form $\chi= a_{tide}\cos(\omega_{tide}t - \varpi)$ for $\omega$ the tidal frequency, $\varpi=\varpi(\boldsymbol{x})$ the relative offset, and $a_{tide}=a_{tide}(\boldsymbol{x})$ the corresponding amplitude of each partcular constituent, each of which determined using the Eastern Pacific Tidal Database discussed in \cite{tidal}.  The remaining variables are quenched by the open ocean boundary.

Finally, the land/beach free boundary at Freeport, TX are taken to satisfy:  \[\begin{aligned}&\quad\qquad\qquad \zeta_{b,1}|_{\partial\Omega_{e_{i}}} = \zeta_{\mathscr{W}\mathscr{D}},\quad \boldsymbol{u}_{b,n,1}|_{\partial\Omega_{e_{i}}} =   \boldsymbol{u}_{\mathscr{W}\mathscr{D}}, \quad \boldsymbol{u}_{b,\tau,1}|_{\partial\Omega_{e_{i}}}=   \boldsymbol{u}_{\mathscr{W}\mathscr{D}} \quad\mathrm{and} \quad \iota_{b,1}|_{\partial\Omega_{e_{i}}} = 0,\end{aligned}\] where $(\cdot)_{\mathscr{W}\mathscr{D}}$ corresponds to the free surface and velocity components are determined by the sophisticated wetting and drying treatment presented in \cite{BKWD}, while the chemical constituents are treated as the inert constituent is in \textsection{4.1}.  

The wetting and drying treatment is beyond the scope of this paper, and we refer the reader to \cite{BKWD} for more details.  It should however be noted that here, elements experiencing ``wetting and drying'' are restricted to the $p=1$ case, as the treatment relies on a re-weighting of the slopes of lines (see \cite{BKWD}) in the case of a linear basis.  This algorithm would require a nontrivial reworking to higher order $p$ to fully achieve a homogeneous $p>1$ solution.  As such, the $p$-enrichment scheme is particularly well suited for just such a model, where one can readily $p$-enrich on interior elements by adapting within $p\in\{1,\ldots,p_{\max}\}$.

The numerical experiments show interesting behavior here.  That is, the $p$ level shown in Figure \ref{fig:shortbrazos} follows the dynamics in the domain, where here we have used the high stability Type II fixed tolerance $p$-enrichment scheme.  We see that along the edge of the tide, at the river inlet and at the boundary of the whirlpool that the $p$ level drops to $p_{\min}$ while elsewhere it adapts to the local perturbation in the solution, introducing a fair amount of additional structure into the polynomial basis.  

In the phosphate neutralization study the $p=1$ cases were run at $\Delta t = 2$ seconds, while the adapting cases were run at $\Delta t = 1$ second.  Some of the results are shown in Figure \ref{fig:freeportfull}.  Generally, we find that when the chemical reaction is active via (\ref{mineraldeposits}) the amount of phosphates present in the Freeport coastal regime is reduced to 10\% that of the solution when the chemistry is turned off.  Interestingly, the addition of the whirlpool mixer (by which here we simply mean the vortical velocity field) does not appreciably improve the extent of the reaction, though it does have the effect of localizing the constituents to the mouth of the estuary and keeping the tidal fluxes from washing the constituents up/down the coast.  The adapting case in Figure \ref{fig:freeportfull} shows very nice behavior here, giving sharper profiles along the edges of the contaminant flow in comparison to the $p=1$ case, and indicates that the model is worthy of further study as a possible route to neutralizing eutrophication from fertilizer contaminants in gulf estuarine flows.  

\section{\texorpdfstring{\protect\centering $\S 5$ Conclusion}{\S 5 Conclusion}}

We have presented a family of $p$-enrichment schemes designed for highly generalizable, computationally efficient, nonlinear free boundary type problems.  This family of schemes was categorized into two classes.  

The first class are the fixed tolerance schemes.  We have found that these schemes are particularly well-suited for problems whose ``energy'' is tightly bounded from both above and below, since the fixed global tolerance otherwise is highly localized in either space or time.  Moreover, the Type I schemes are attractive to many model applications which desire to emphasize particular ``high energy regions'' of the domain for enrichment, while the Type II schemes are appealing to application models where stability and computational efficiency trump the former concerns.

The second class are the dioristic enrichment schemes.  These schemes address the most obvious drawback of the fixed tolerance schemes, in that they adapt locally in time to the global spatial bounds of the regularity estimator.  These schemes are well-suited for contexts which are complementary to those of the fixed tolerance schemes: namely, contexts where the ``energy'' varies substantially in both $x$ and $t$.  

We then studied a pair of model problems.  The first, a contaminant declinature model, was used to study the $p$-convergence of the solution, the accuracy and robustness behavior of the enrichment schemes, and the effect of weak entropy boundary layers.  The physics of the model suggests that imparting mechanical wave energy might provide a simple way of diverting inert contaminants near important coastal regions.  The second model was an estuary eutrophication study of a reactive multicomponent flow regime, where fertilizer runoff into a dead zone was counteracted by way of a neutralization reaction at the mouth of the estuary, and suggests that such a preliminary model might be able to provide a remediation mechanism in these areas.  

Future work is largely aimed towards model verification by way of employing the $p$-enrichment schemes to efficiently improve the accuracy of large applications models, such as hurricane storm surge.  We also note that a very beautiful extension of the coupled chemical model from \textsection{4.2} is presented in \cite{Alvarez}, which includes the additional parameters of volatilization and sorption rates of chemicals in the atmosphere and suspended sediment, respectively, with the water.  Studying the effects of these parameters on the eutrophication model, for example, might provide a more realistic understanding, as would verifying the hydraulicity and ecological impact of (\ref{mineraldeposits}) in the context of sea water. 

\section{\texorpdfstring{\protect\centering $\S 6$ Acknowledgements}{\S 6 Acknowledgements}}

The authors would also like to aknowledge the support of the National Science Foundation grants OCI-0749075 and OCI-0746232.

\def\cprime{$'$} \def\cprime{$'$}
  \def\polhk#1{\setbox0=\hbox{#1}{\ooalign{\hidewidth
  \lower1.5ex\hbox{`}\hidewidth\crcr\unhbox0}}}
  \def\polhk#1{\setbox0=\hbox{#1}{\ooalign{\hidewidth
  \lower1.5ex\hbox{`}\hidewidth\crcr\unhbox0}}}
  \def\polhk#1{\setbox0=\hbox{#1}{\ooalign{\hidewidth
  \lower1.5ex\hbox{`}\hidewidth\crcr\unhbox0}}} \def\cprime{$'$}
  \def\cprime{$'$} \def\polhk#1{\setbox0=\hbox{#1}{\ooalign{\hidewidth
  \lower1.5ex\hbox{`}\hidewidth\crcr\unhbox0}}}
  \def\polhk#1{\setbox0=\hbox{#1}{\ooalign{\hidewidth
  \lower1.5ex\hbox{`}\hidewidth\crcr\unhbox0}}}
  \def\polhk#1{\setbox0=\hbox{#1}{\ooalign{\hidewidth
  \lower1.5ex\hbox{`}\hidewidth\crcr\unhbox0}}} \def\cprime{$'$}
  \def\cprime{$'$} \def\polhk#1{\setbox0=\hbox{#1}{\ooalign{\hidewidth
  \lower1.5ex\hbox{`}\hidewidth\crcr\unhbox0}}}
  \def\polhk#1{\setbox0=\hbox{#1}{\ooalign{\hidewidth
  \lower1.5ex\hbox{`}\hidewidth\crcr\unhbox0}}}
  \def\polhk#1{\setbox0=\hbox{#1}{\ooalign{\hidewidth
  \lower1.5ex\hbox{`}\hidewidth\crcr\unhbox0}}} \def\cprime{$'$}
  \def\cprime{$'$} \def\polhk#1{\setbox0=\hbox{#1}{\ooalign{\hidewidth
  \lower1.5ex\hbox{`}\hidewidth\crcr\unhbox0}}}
  \def\polhk#1{\setbox0=\hbox{#1}{\ooalign{\hidewidth
  \lower1.5ex\hbox{`}\hidewidth\crcr\unhbox0}}}
  \def\polhk#1{\setbox0=\hbox{#1}{\ooalign{\hidewidth
  \lower1.5ex\hbox{`}\hidewidth\crcr\unhbox0}}}


\begin{thebibliography}{50}
\providecommand{\natexlab}[1]{#1}
\providecommand{\url}[1]{\texttt{#1}}
\expandafter\ifx\csname urlstyle\endcsname\relax
  \providecommand{\doi}[1]{doi: #1}\else
  \providecommand{\doi}{doi: \begingroup \urlstyle{rm}\Url}\fi

\bibitem[Aizinger and Dawson(2002)]{Aizinger1}
V.~Aizinger and C.~Dawson.
\newblock A discontinuous galerkin method for two-dimensional flow and
  transport in shallow water.
\newblock \emph{Advances in Water Resources}, 25\penalty0 (1):\penalty0 67 --
  84, 2002.
\newblock ISSN 0309-1708.
\newblock \doi{DOI: 10.1016/S0309-1708(01)00019-7}.
\newblock URL
  \url{http://www.sciencedirect.com/science/article/B6VCF-44PK3KB-5/2/51beaaea%
1191c299bcd3a0d40beca43d}.

\bibitem[Alvarez-V\'{a}zquez et~al.(2009)Alvarez-V\'{a}zquez, Mart\'{\i}nez,
  V\'{a}zquez-M\'{e}ndez, and Vilar]{Alvarez}
L.~J. Alvarez-V\'{a}zquez, A.~Mart\'{\i}nez, M.~E. V\'{a}zquez-M\'{e}ndez, and
  M.~A. Vilar.
\newblock An application of optimal control theory to river pollution
  remediation.
\newblock \emph{Appl. Numer. Math.}, 59:\penalty0 845--858, May 2009.
\newblock ISSN 0168-9274.
\newblock \doi{10.1016/j.apnum.2008.03.027}.
\newblock URL \url{http://portal.acm.org/citation.cfm?id=1517854.1518083}.

\bibitem[Arnold et~al.(2000)Arnold, Brezzi, Cockburn, and Marini]{ABCM}
D.N. Arnold, F.~Brezzi, B.~Cockburn, and D.~Marini.
\newblock Discontinuous {G}alerkin methods for elliptic problems.
\newblock In \emph{Discontinuous Galerkin methods (Newport, RI, 1999)},
  volume~11 of \emph{Lect. Notes Comput. Sci. Eng.}, pages 89--101. Springer,
  Berlin, 2000.

\bibitem[Bey et~al.(1995)Bey, Patra, and Oden]{Bey2}
K.~S. Bey, A.~Patra, and J.~T. Oden.
\newblock {$hp$}-version discontinuous {G}alerkin methods for hyperbolic
  conservation laws: a parallel adaptive strategy.
\newblock \emph{Internat. J. Numer. Methods Engrg.}, 38\penalty0 (22):\penalty0
  3889--3908, 1995.
\newblock ISSN 0029-5981.
\newblock \doi{10.1002/nme.1620382209}.
\newblock URL \url{http://dx.doi.org/10.1002/nme.1620382209}.

\bibitem[Bey et~al.(1996)Bey, Oden, and Patra]{Bey1}
K.S. Bey, J.~T. Oden, and A.~Patra.
\newblock A parallel {$hp$}-adaptive discontinuous {G}alerkin method for
  hyperbolic conservation laws.
\newblock \emph{Appl. Numer. Math.}, 20\penalty0 (4):\penalty0 321--336, 1996.
\newblock ISSN 0168-9274.
\newblock \doi{10.1016/0168-9274(95)00101-8}.
\newblock URL \url{http://dx.doi.org/10.1016/0168-9274(95)00101-8}.
\newblock Adaptive mesh refinement methods for CFD applications (Atlanta, GA,
  1994).

\bibitem[Bresch et~al.(2007)Bresch, Desjardins, and M{\'e}tivier]{BD2}
D.~Bresch, B.~Desjardins, and G.~M{\'e}tivier.
\newblock Recent mathematical results and open problems about shallow water
  equations.
\newblock In \emph{Analysis and simulation of fluid dynamics}, Adv. Math. Fluid
  Mech., pages 15--31. Birkh\"auser, Basel, 2007.

\bibitem[Brown and Fulmer(1991)]{BrownP}
P.W. Brown and M.~Fulmer.
\newblock Kinetics of hydroxyapatite formation at low temperature.
\newblock \emph{Journal of the American Ceramic Society}, 74\penalty0
  (5):\penalty0 934--940, 1991.
\newblock ISSN 1551-2916.
\newblock \doi{10.1111/j.1151-2916.1991.tb04324.x}.
\newblock URL \url{http://dx.doi.org/10.1111/j.1151-2916.1991.tb04324.x}.

\bibitem[Bunya et~al.(2009)Bunya, Kubatko, Westerink, and Dawson]{BKWD}
S.~Bunya, E.J. Kubatko, J.~J. Westerink, and C.~Dawson.
\newblock A wetting and drying treatment for the {R}unge-{K}utta discontinuous
  {G}alerkin solution to the shallow water equations.
\newblock \emph{Comput. Methods Appl. Mech. Engrg.}, 198\penalty0
  (17-20):\penalty0 1548--1562, 2009.
\newblock ISSN 0045-7825.
\newblock \doi{10.1016/j.cma.2009.01.008}.
\newblock URL \url{http://dx.doi.org/10.1016/j.cma.2009.01.008}.

\bibitem[Burbeau and Sagaut(2005)]{BurbeauS}
A.~Burbeau and P.~Sagaut.
\newblock A dynamic {$p$}-adaptive discontinuous {G}alerkin method for viscous
  flow with shocks.
\newblock \emph{Comput. \& Fluids}, 34\penalty0 (4-5):\penalty0 401--417, 2005.
\newblock ISSN 0045-7930.
\newblock \doi{10.1016/j.compfluid.2003.04.002}.
\newblock URL \url{http://dx.doi.org/10.1016/j.compfluid.2003.04.002}.

\bibitem[Chan et~al.(2010)Chan, Demkowicz, Moser, and Roberts]{dem2}
J.~Chan, L.~Demkowicz, R.~Moser, and N.~Roberts.
\newblock A new discontinuous {P}etrov-{G}alerkin method with optimal test
  functions. part v: Solution of 1d burgers’ and navier-stokes equations.
\newblock page~34, 2010.
\newblock URL \url{http://www.ices.utexas.edu/media/reports/2010/1025.pdf}.

\bibitem[Cheng et~al.(2000)Cheng, Yeh, and Cheng]{Cheng}
H.~P. Cheng, G.~T. Yeh, and J.~R. Cheng.
\newblock A numerical model simulating reactive transport in shallow water
  domains: model development and demonstrative applications.
\newblock \emph{Advances in Environmental Research}, 4\penalty0 (3):\penalty0
  187 -- 209, 2000.
\newblock ISSN 1093-0191.
\newblock \doi{DOI: 10.1016/S1093-0191(00)00015-0}.
\newblock URL
  \url{http://www.sciencedirect.com/science/article/B6W75-412RWTW-2/2/3e18b73a%
2485c63d4435e1a37124e8e2}.

\bibitem[Dawson and Aizinger({2002})]{Aizinger2}
C.~Dawson and V.~Aizinger.
\newblock {The local discontinuous galerkin method for advection-diffusion
  equations arising in groundwater and surface water applications}.
\newblock In {Chadam, J and Cunningham, A and Ewing, RE and Ortoleva, P and
  Wheller, MF}, editor, \emph{Resource Recovery, Confinement, and Remediation
  of Environmental Hazards}, volume {131} of \emph{IMA Volumes in Mathematics
  and its Applications}, pages {231--245}, {233 Spring Street, New York, NY
  10013, United States}, {2002}. {Inst Math \& Applicat}, Springer.
\newblock ISBN {0-387-95506-2}.
\newblock {Workshop on Resource Recovery, Minneapolis, MN, Jan 15-19, 2000}.

\bibitem[Dawson et~al.(2010)Dawson, Kubatko, Westerink, Trahan, Mirabito,
  Michoski, and Panda]{Dawson2010}
C.~Dawson, E.J. Kubatko, J.J. Westerink, C.~Trahan, C.~Mirabito, C.~Michoski,
  and N.~Panda.
\newblock Discontinuous galerkin methods for modeling hurricane storm surge.
\newblock \emph{Advances in Water Resources}, In Press, Corrected
  Proof:\penalty0 --, 2010.
\newblock ISSN 0309-1708.
\newblock \doi{DOI: 10.1016/j.advwatres.2010.11.004}.
\newblock URL
  \url{http://www.sciencedirect.com/science/article/B6VCF-51JXFRT-1/2/e33be9cc%
dde82554ef9018938100d12f}.

\bibitem[Dedner and Ohlberger(2008)]{Dedner}
A.~Dedner and M.~Ohlberger.
\newblock A new {$hp$}-adaptive {DG} scheme for conservation laws based on
  error control.
\newblock In \emph{Hyperbolic problems: theory, numerics, applications}, pages
  187--198. Springer, Berlin, 2008.
\newblock \doi{10.1007/978-3-540-75712-2_15}.
\newblock URL \url{http://dx.doi.org/10.1007/978-3-540-75712-2_15}.

\bibitem[Demkowicz(2007)]{Dem}
L.~Demkowicz.
\newblock \emph{Computing with {$hp$}-adaptive finite elements. {V}ol. 1}.
\newblock Chapman \& Hall/CRC Applied Mathematics and Nonlinear Science Series.
  Chapman \& Hall/CRC, Boca Raton, FL, 2007.
\newblock ISBN 978-1-58488-671-6; 1-58488-671-4.
\newblock One and two dimensional elliptic and Maxwell problems, With 1 CD-ROM
  (UNIX).

\bibitem[Demkowicz et~al.(2008)Demkowicz, Kurtz, Pardo, Paszy{\'n}ski,
  Rachowicz, and Zdunek]{Demk2}
L.~Demkowicz, J.~Kurtz, D.~Pardo, M.~Paszy{\'n}ski, W.~Rachowicz, and
  A.~Zdunek.
\newblock \emph{Computing with {$hp$}-adaptive finite elements. {V}ol. 2}.
\newblock Chapman \& Hall/CRC Applied Mathematics and Nonlinear Science Series.
  Chapman \& Hall/CRC, Boca Raton, FL, 2008.
\newblock ISBN 978-1-58488-672-3; 1-58488-672-2.
\newblock Frontiers: three dimensional elliptic and Maxwell problems with
  applications.

\bibitem[Feistauer et~al.(2003)Feistauer, Felcman, and Stra{\v{s}}kraba]{FFS}
M.~Feistauer, J.~Felcman, and I.~Stra{\v{s}}kraba.
\newblock \emph{Mathematical and computational methods for compressible flow}.
\newblock Numerical mathematics and scientific computation. Oxford University
  Press, 2003.
\newblock ISBN 0-19-850588-4.

\bibitem[Giovangigli(1999)]{Giovangigli3}
V.~Giovangigli.
\newblock \emph{Multicomponent flow modeling}.
\newblock Modeling and Simulation in Science, Engineering and Technology.
  Birkh\"auser Boston Inc., Boston, MA, 1999.
\newblock ISBN 0-8176-4048-7.

\bibitem[Heemink(1993)]{Heemink}
A.~W. Heemink.
\newblock Tidally averaged models for dispersion in shallow water.
\newblock \emph{Water Resour. Res.}, 29\penalty0 (3):\penalty0 607--617, 1993.

\bibitem[Hirschfelder et~al.(1954)Hirschfelder, Curtiss, and
  Bird]{Hirschfelder}
J.O. Hirschfelder, C.F. Curtiss, and R.B. Bird.
\newblock \emph{The Molecular Theory of Gases and Liquids}.
\newblock Structure of Matter Series. Wiley-Interscience, Revised, New York,
  1954.
\newblock ISBN 0-47-1400653, 978-0471400653.

\bibitem[Irvine et~al.(2006)Irvine, Mann, and Short]{IMS}
G.V. Irvine, D.H. Mann, and J.W. Short.
\newblock Persistence of 10-year old exxon valdez oil on gulf of alaska
  beaches: The importance of boulder-armoring.
\newblock \emph{Marine Pollution Bulletin}, 52\penalty0 (9):\penalty0 1011 --
  1022, 2006.
\newblock ISSN 0025-326X.
\newblock \doi{DOI: 10.1016/j.marpolbul.2006.01.005}.
\newblock URL
  \url{http://www.sciencedirect.com/science/article/B6V6N-4JF8JN0-3/2/f7df9b25%
43cd0bee96550765e9ab5f5d}.

\bibitem[Johnson et~al.(1965)Johnson, Rawson, and Smith]{JRS}
S.L. Johnson, J.~Rawson, and R.E. Smith.
\newblock Characteristics of tide-affected flow in the brazos river near
  freeport, texas.
\newblock \emph{Texas Water Development Board}, 69, 1965.

\bibitem[Kanazawa et~al.(2000)Kanazawa, Monma, and Moriyoshi]{Kana}
T.~Kanazawa, H.~Monma, and Y.~Moriyoshi.
\newblock Reaction chemistry of hydroxyapatite: Formation and decomposition.
\newblock \emph{Proceedings of the Estonian Academy of Sciences, Chemistry},
  49\penalty0 (1):\penalty0 19--28, 2000.

\bibitem[Katsuki and Komarneni(1998)]{KatKom}
H.~Katsuki and S.~Komarneni.
\newblock Porous hydroxyapatite monoliths from gypsum waste.
\newblock \emph{J. Mater. Chem.}, 8\penalty0 (12):\penalty0 2803, 1998.
\newblock ISSN 0959-9428.

\bibitem[Kubatko et~al.(2006)Kubatko, Westerink, and Dawson]{KubatkoWD}
E.J. Kubatko, J.J. Westerink, and C.~Dawson.
\newblock hp discontinuous galerkin methods for advection dominated problems in
  shallow water flow.
\newblock \emph{Computer Methods in Applied Mechanics and Engineering},
  196\penalty0 (1-3):\penalty0 437 -- 451, 2006.
\newblock ISSN 0045-7825.
\newblock \doi{DOI: 10.1016/j.cma.2006.05.002}.
\newblock URL
  \url{http://www.sciencedirect.com/science/article/B6V29-4M1CYTM-1/2/6c45c85d%
20d17690046881a795b0b04d}.

\bibitem[Kubatko et~al.(2007)Kubatko, Westerink, and Dawson]{KWD}
E.J. Kubatko, J.J. Westerink, and C.~Dawson.
\newblock Semi discrete discontinuous {G}alerkin methods and
  stage-exceeding-order, strong-stability-preserving {R}unge-{K}utta time
  discretizations.
\newblock \emph{J. Comput. Phys.}, 222\penalty0 (2):\penalty0 832--848, 2007.
\newblock ISSN 0021-9991.
\newblock \doi{10.1016/j.jcp.2006.08.005}.
\newblock URL \url{http://dx.doi.org/10.1016/j.jcp.2006.08.005}.

\bibitem[Kubatko et~al.(2008)Kubatko, Dawson, and Westerink]{KDW}
E.J. Kubatko, C.~Dawson, and J.J. Westerink.
\newblock Time step restrictions for {R}unge-{K}utta discontinuous {G}alerkin
  methods on triangular grids.
\newblock \emph{J. Comput. Phys.}, 227\penalty0 (23):\penalty0 9697--9710,
  2008.
\newblock ISSN 0021-9991.
\newblock \doi{10.1016/j.jcp.2008.07.026}.
\newblock URL \url{http://dx.doi.org/10.1016/j.jcp.2008.07.026}.

\bibitem[Kubatko et~al.(2009{\natexlab{a}})Kubatko, Bunya, Dawson, and
  Westerink]{KubBDW}
E.J. Kubatko, S.~Bunya, C.~Dawson, and J.J. Westerink.
\newblock Dynamic p-adaptive runge-kutta discontinuous galerkin methods for the
  shallow water equations.
\newblock \emph{Computer Methods in Applied Mechanics and Engineering},
  198\penalty0 (21-26):\penalty0 1766 -- 1774, 2009{\natexlab{a}}.
\newblock ISSN 0045-7825.
\newblock \doi{DOI: 10.1016/j.cma.2009.01.007}.
\newblock URL
  \url{http://www.sciencedirect.com/science/article/B6V29-4VDY7X4-1/2/36e49328%
fea4e4f751d689510b7e3b3f}.
\newblock Advances in Simulation-Based Engineering Sciences - Honoring J.
  Tinsley Oden.

\bibitem[Kubatko et~al.(2009{\natexlab{b}})Kubatko, Bunya, Dawson, Westerink,
  and Mirabito]{KBDWM}
E.J. Kubatko, S.~Bunya, C.~Dawson, J.J. Westerink, and C.~Mirabito.
\newblock A performance comparison of continuous and discontinuous finite
  element shallow water models.
\newblock \emph{J. Sci. Comput.}, 40\penalty0 (1-3):\penalty0 315--339,
  2009{\natexlab{b}}.
\newblock ISSN 0885-7474.
\newblock \doi{10.1007/s10915-009-9268-2}.
\newblock URL \url{http://dx.doi.org/10.1007/s10915-009-9268-2}.

\bibitem[Lipscomb and Ringler(2005)]{LipRing}
W.~H. Lipscomb and T.~D. Ringler.
\newblock An incremental remapping transport scheme on a spherical geodesic
  grid.
\newblock \emph{Monthly Weather Review}, 133\penalty0 (8):\penalty0 2335--2350,
  2005.
\newblock \doi{10.1175/MWR2983.1}.
\newblock URL \url{http://journals.ametsoc.org/doi/abs/10.1175/MWR2983.1}.

\bibitem[Luettich et~al.({1998})Luettich, Hench, Williams, Blanton, and
  Werner]{Luettich}
RA~Luettich, JL~Hench, CD~Williams, BO~Blanton, and FE~Werner.
\newblock {Tidal circulation and larval transport through a barrier island
  inlet}.
\newblock In {Spaulding, ML and Blumberg, AF}, editor, \emph{{Estuarine and
  Coastal Modeling}}, pages {849--863}, {United Engineering Center, 345 E 47th
  St, New York, NY 10017-2398 USA}, {1998}. {Amer Soc Civil Engineers}.
\newblock ISBN {0-7844-0350-3}.
\newblock {5th International Conference on Estuarine and Coastal Modeling,
  Alexandria, VA, Oct 22-24, 1997}.

\bibitem[MacQuarrie and Sudicky(2001)]{MacQuarrie}
K.T.B. MacQuarrie and E.A. Sudicky.
\newblock Multicomponent simulation of wastewater-derived nitrogen and carbon
  in shallow unconfined aquifers: I. model formulation and performance.
\newblock \emph{Journal of Contaminant Hydrology}, 47\penalty0 (1):\penalty0 53
  -- 84, 2001.
\newblock ISSN 0169-7722.
\newblock \doi{DOI: 10.1016/S0169-7722(00)00137-6}.
\newblock URL
  \url{http://www.sciencedirect.com/science/article/B6V94-41WJBWM-3/2/5651b451%
0f03c9e5e072df5d9fff5d1f}.

\bibitem[Mark et~al.(2004)Mark, Spargo, Westerink, and Leuttich]{tidal}
D.J. Mark, E.A. Spargo, J.J. Westerink, and R.A. Leuttich.
\newblock {ENPAC} 2003: {A} {T}idal {C}onstituent {D}atabase for {E}astern
  {N}orth {P}acific {O}cean.
\newblock \emph{U.S Army Corps of Engineers, Report}, pages 1--191, 2004.
\newblock URL \url{http://www.unc.edu/ims/ccats/tides/ENPAC_2003_report.pdf}.

\bibitem[Martin(2007)]{Martin}
S.~Martin.
\newblock First order quasilinear equations with boundary conditions in the
  {$L\sp \infty$} framework.
\newblock \emph{J. Differential Equations}, 236\penalty0 (2):\penalty0
  375--406, 2007.
\newblock ISSN 0022-0396.

\bibitem[Michoski and Schmitz(2011)]{MS}
C.~Michoski and P.G. Schmitz.
\newblock Multiscale $hp$-adaptive {C}hemical {R}eacters {II}: {L}abile
  {C}atalytic {R}eactors.
\newblock \emph{preprint}, 2011.

\bibitem[Michoski et~al.(2010{\natexlab{a}})Michoski, Dawson, Kubatko,
  Mirabito, Westerink, and Wirasaet]{Mich2}
C.~Michoski, C.~Dawson, E.J. Kubatko, C.~Mirabito, J.J. Westerink, and
  D.~Wirasaet.
\newblock Adaptive hierarchic transformations for dynamically p-enriched
  slope-limiting over discontinuous {G}alerkin systems of generalized
  equations.
\newblock \emph{preprint}, \emph{submitted}, 2010{\natexlab{a}}.

\bibitem[Michoski et~al.(2010{\natexlab{b}})Michoski, Evans, Schmitz, and
  Vasseur]{MESV}
C.~Michoski, J.A. Evans, P.G. Schmitz, and A.~Vasseur.
\newblock A discontinuous {G}alerkin method for viscous compressible
  multifluids.
\newblock \emph{J. Comput. Phys.}, 229\penalty0 (6):\penalty0 2249--2266,
  2010{\natexlab{b}}.
\newblock ISSN 0021-9991.
\newblock \doi{10.1016/j.jcp.2009.11.033}.
\newblock URL \url{http://dx.doi.org/10.1016/j.jcp.2009.11.033}.

\bibitem[Michoski et~al.(2011)Michoski, Evans, and Schmitz]{MES}
C.~Michoski, J.A. Evans, and P.G. Schmitz.
\newblock Multiscale $hp$-adaptive {C}hemical {R}eacters {I}: {Q}uiescent
  {R}eactors.
\newblock \emph{preprint}, \emph{In Review}, 2011.

\bibitem[Mirabito et~al.(2011)Mirabito, Dawson, Kubatko, Westerink, and
  Bunya]{Mirabito}
C.~Mirabito, C.~Dawson, E.~J. Kubatko, J.~J. Westerink, and S.~Bunya.
\newblock Implementation of a discontinuous {G}alerkin morphological model on
  two-dimensional unstructured meshes.
\newblock \emph{Comput.\ Methods Appl.\ Mech.\ Engrg.}, 200\penalty0
  (1--4):\penalty0 189--207, January 2011.

\bibitem[\"{O}rs(1999)]{Ors}
H.~\"{O}rs.
\newblock Shallow water model for the bosphorus current.
\newblock In Sedat Biringen, Haluk Örs, Akin Tezel, and Joel Ferziger,
  editors, \emph{Industrial and Environmental Applications of Direct and
  Large-Eddy Simulation}, volume 529 of \emph{Lecture Notes in Physics}, pages
  241--247. Springer Berlin / Heidelberg, 1999.
\newblock URL \url{http://dx.doi.org/10.1007/BFb0106107}.
\newblock 10.1007/BFb0106107.

\bibitem[\"{O}rs and Yilmaz({2004})]{Ors2}
H~\"{O}rs and SL~Yilmaz.
\newblock {A stochastic approach to modeling of oil pollution}.
\newblock \emph{{ENERGY SOURCES}}, {26}\penalty0 ({9}):\penalty0 {879--884},
  {JUL 16} {2004}.
\newblock ISSN {0090-8312}.
\newblock \doi{{10.1080/00908310490465920}}.

\bibitem[Palaniappan et~al.(2008)Palaniappan, Miller, and Haber]{PMH}
J.~Palaniappan, S.T. Miller, and R.B. Haber.
\newblock Sub-cell shock capturing and spacetime discontinuity tracking for
  nonlinear conservation laws.
\newblock \emph{Internat. J. Numer. Methods Fluids}, 57\penalty0 (9):\penalty0
  1115--1135, 2008.
\newblock ISSN 0271-2091.
\newblock \doi{10.1002/fld.1850}.
\newblock URL \url{http://dx.doi.org/10.1002/fld.1850}.

\bibitem[Persson and Peraire(2006)]{PPe}
P.P. Persson and J.~Peraire.
\newblock Sub-cell shock capturing for discontinuous galerkin methods.
\newblock \emph{Forty-fourth AIAA Aerospace Sciences Meeting and Exhibit, Reno,
  NV, U.S.A.}, Online:\penalty0 5--–18, 2006.

\bibitem[Ruuth(2006)]{Ruuth}
S.J. Ruuth.
\newblock Global optimization of explicit strong-stability-preserving
  {R}unge-{K}utta methods.
\newblock \emph{Math. Comp.}, 75\penalty0 (253):\penalty0 183--207
  (electronic), 2006.
\newblock ISSN 0025-5718.
\newblock \doi{10.1090/S0025-5718-05-01772-2}.
\newblock URL \url{http://dx.doi.org/10.1090/S0025-5718-05-01772-2}.

\bibitem[Sch{\"o}tzau and Schwab(2000)]{Schotzau}
D.~Sch{\"o}tzau and C.~Schwab.
\newblock An {$hp$} a priori error analysis of the {DG} time-stepping method
  for initial value problems.
\newblock \emph{Calcolo}, 37\penalty0 (4):\penalty0 207--232, 2000.
\newblock ISSN 0008-0624.
\newblock \doi{10.1007/s100920070002}.
\newblock URL \url{http://dx.doi.org/10.1007/s100920070002}.

\bibitem[Shu and Osher(1988)]{SO}
C.-W. Shu and S.~Osher.
\newblock Efficient implementation of essentially nonoscillatory
  shock-capturing schemes.
\newblock \emph{J. Comput. Phys.}, 77\penalty0 (2):\penalty0 439--471, 1988.
\newblock ISSN 0021-9991.

\bibitem[Solonnikov and Tani(1992)]{ST}
V.~A. Solonnikov and A.~Tani.
\newblock Evolution free boundary problem for equations of motion of viscous
  compressible barotropic liquid.
\newblock In \emph{The Navier-Stokes equations II---theory and numerical
  methods (Oberwolfach, 1991)}, volume 1530 of \emph{Lecture Notes in Math.},
  pages 30--55. Springer, Berlin, 1992.

\bibitem[Sylvan et~al.(2006)Sylvan, Dortch, Nelson, Maier~Brown, Morrison, and
  Ammerman]{SDNMMA}
J.B. Sylvan, Q.~Dortch, D.M. Nelson, A.F. Maier~Brown, W.~Morrison, and J.W.
  Ammerman.
\newblock Phosphorus limits phytoplankton growth on the {L}ouisiana shelf
  during the period of hypoxia formation.
\newblock \emph{Environmental Science \& Technology}, 40\penalty0
  (24):\penalty0 7548--7553, 2006.
\newblock \doi{10.1021/es061417t}.
\newblock URL \url{http://pubs.acs.org/doi/abs/10.1021/es061417t}.

\bibitem[Vreugdenhil(1998)]{Vre}
C.B. Vreugdenhil.
\newblock \emph{Numerical Methods for Shallow-Watr Flow}, volume 1st Edition
  reprint.
\newblock Kluwer Academic Publishers, Netherlands, 1998.
\newblock ISBN 0-792-33164-8.

\bibitem[Wang and Mavriplis(2009)]{WanMa}
L.~Wang and D.J. Mavriplis.
\newblock Adjoint-based {$h$}-{$p$} adaptive discontinuous {G}alerkin methods
  for the 2{D} compressible {E}uler equations.
\newblock \emph{J. Comput. Phys.}, 228\penalty0 (20):\penalty0 7643--7661,
  2009.
\newblock ISSN 0021-9991.
\newblock \doi{10.1016/j.jcp.2009.07.012}.
\newblock URL \url{http://dx.doi.org/10.1016/j.jcp.2009.07.012}.

\end{thebibliography}
\end{document}